
\input phyzzx

\def\IMPA#1{{\sl Int. J. Mod. Phys. {\bf A#1}}}
\def\IMPB#1{{\sl Int. J. Mod. Phys. {\bf B#1}}}

\def\JMP#1{{\sl J.\ Math. \ Phys.\ {\bf #1}}}

\def\JPA#1{{\sl J.\ Phys.\ {\bf A#1}}}

\def\MPL#1{{\sl Mod.\ Phys.\ Lett. \ {\bf #1}}}

\def\NPB#1{{\sl Nucl.\ Phys.\ {\bf B#1}}}

\def\PLB#1{{\sl Phys.\ Lett.\ {\bf #1B}}}

\def\PR#1{{\sl Phys.\ Rev.\ {\bf #1}}}

\def\PRB#1{{\sl Phys.\ Rev.\   {\bf B #1}}}
\def\PRL#1{{\sl Phys.\ Rev.\ Lett.\ {\bf #1}}}

\def\PTP#1{{\sl Prog.\ Theor.\ Phys.\ {\bf #1}}}

\def\ZPB#1{{\sl Z.\ Phys.\ {\bf B}\ {\bf #1}}}


\def\nxl{\hfill\break}


\def\D{{\cal D}}
\def\H{{\cal H}}                            
\def\Z{{\cal Z}}                            

\def\a{\alpha}

\def\b{\beta}
\def\g{\gamma}

\def\e{\epsilon}

\def\l{\lambda}

\def\s{\sigma}
\def\om{\omega}
\def\t{\theta}

\def\Th{\Theta}

\def\o{\over}

\def\bold#1{\setbox0=\hbox{$#1$}
     \kern-.025em\copy0\kern-\wd0
     \kern.05em\copy0\kern-\wd0
     \kern-.025em\raise.0433em\box0 }
\def\lowmp{\lower.11em\hbox{${\scriptstyle\mp}$}}

\def\Im{{\rm Im\,}}
\def\Re{{\rm Re\,}}

\def\frac#1#2{{\textstyle{
 #1 \over #2 }}}                            


\def\ZZ{{\rm Z \!\! Z}}                       
\def\1{{\rm 1 \!\!\, l}}                        
%
\def\partder#1#2{{\partial #1\over\partial #2}}

%
%


\Pubnum={$\rm IFUM 477/FT; \; LPTHE \; 94-28 ; \;  hep-th/9407117;
\quad {\rm July \; 1994}$}
\date={}
\titlepage
\title{UNIFIED
APPROACH TO THERMODYNAMIC BETHE ANSATZ  AND FINITE SIZE CORRECTIONS FOR
LATTICE MODELS AND FIELD THEORIES}
\author{ C. Destri }
\address{ Dipartimento di Fisica, Universit\`a di Milano
          and INFN, sezione di Milano
     \foot{mail address: \nxl
           Dipartimento di Fisica \nxl
           Via Celoria 16, 20133 Milano ITALIA }}

\author{ H.J. de Vega }
\address{ Laboratoire de Physique Th\'eorique et Hautes Energies
     \foot{Laboratoire Associ\'e au CNRS UA 280}$\!$, Paris
     \foot{mail address: \nxl
           L.P.T.H.E., Tour 16 $1^{\rm er}$ \'etage, Universit\'e Paris VI,\nxl
           4 Place Jussieu, 75252, Paris cedex 05, FRANCE }}

\vfil
\abstract
We present a unified approach to the Thermodynamic  Bethe Ansatz (TBA) for
magnetic chains and field theories that includes the finite size (and
zero temperature) calculations for lattice BA models.
In all cases, the free energy follows by quadratures from the solution
of a {\bf single} non-linear integral equation (NLIE). [A system of
NLIE appears for nested BA].

We derive the NLIE for: a) the six-vertex model with twisted
boundary conditions; b) the XXZ chain in an external magnetic field
$h_z$  and c) the sine-Gordon-massive Thirring model (sG-mT) in a periodic
box of size $\b \equiv 1/T $ using the light-cone approach.
This NLIE is solved by iteration in one
regime (high $T$ in the XXZ chain and low $T$ in the sG-mT model). In
the opposite (conformal) regime, the leading behaviors are
obtained in closed form. Higher corrections can be derived from the
Riemann-Hilbert form of the NLIE that we present.

\medskip

\vskip 1cm

\def\intf{\int_{-\infty}^{+\infty}\!}
\def\bt{{\tilde\beta}}
\def\phip{\phi^\prime}
\def\d{\delta}
\def\fb{{\bar f}}
\def\Im{{\rm Im\,}}
\def\Re{{\rm Re\,}}
\def\til{\tilde}
\def\gt{{\hat\g}}
\def\ht{{\tilde h}}
\REF\yy{C.N. Yang and C.P. Yang, \JMP{10} (1969) 1115.}
\REF\tak{M. Takahashi, \PTP{46} (1971) 401. \nxl
         M. Takahashi and M. Suzuki, \PTP{48} (1972) 2187.}
\REF\gaud{M. Gaudin, \PRL{26} (1971) 1302.}
\REF\albz{Al.B. Zamolodchikov, \NPB{342} (1990) 695.}
\REF\tbam{Al.B. Zamolodchikov, \NPB{358} (1991) 524.\nxl
          P. Christe and M. Martins, \MPL{A5} (1990) 2189. \nxl
          M. Martins, \PRL{65} (1990) 2091.\nxl
          T. Klassen and E. Melzer, \NPB{338} (1990) 485;  \nxl
          \NPB{350} (1991) 635.\nxl
          F. Ravanini, \PLB{282} (1992) 73.\nxl
          V. A. Fateev and Al. B. Zamolodchikov, \PLB{271} (1991) 91.\nxl
          P. Fendley, K. Intriligator  \NPB{372} (1992) 533.\nxl
          F.Ravanini, R. Tateo and A. Valleriani, \IMPA{8} (1993) 1707.}
\REF\nos{C. Destri and H. J. de Vega, \PRL{69} (1992) 2313 .}
\REF\workinprogress{C. Destri and H. J. de Vega, work in progress.}
\REF\klum{A. Klumper, \ZPB{91} (1993) 507.}
\REF\baxt{R.J. Baxter, ``{\it Exactly solved models in Statistical Mechanics"},
Academic Press, 1982. }
\REF\suz{M. Suzuki, \PTP{56} (1976) 1454.}
\REF\japs{J. Suzuki, Y. Akutsu and M. Wadati, {\sl J. Phys. Soc. Jpn.} {\bf 59}
          (1990) 2667.}
\REF\taka{T. Koma, \PTP{81} (1988) 783. \nxl
   M. Takahashi , \PRB{43}, 5788 (1990).}
\REF\ddev{C. Destri and H.J. de Vega, \NPB{290} (1987) 363.}
\REF\fata{L. D. Faddeev and L. A. Takhtadzhyan, {\sl Russian Math. Surveys}
          {\bf 34} (1979) 11.}
\REF\dev{H.J. de Vega, \IMPA{} (1989) 2371; \IMPB{} (1990) 735.}
\REF\devw{H.J. de Vega and F. Woynarovich, \NPB{251} (1985) 439.}
\REF\devk{H.J. de Vega and M. Karowski, \NPB{280} (1987) 225.}
\REF\aus{A. Klumper, M.T. Batchelor and P.A. Pearce, \JPA{24} (1991) 3111.}
\REF\dedev{C. Destri and H.J. de Vega, \NPB{258} (1991) 251.}
\REF\ddvj{C. Destri and H.J. de Vega, \JPA{22} (1989) 1329.}
\REF\ybs{C. Destri and H. J. de Vega, \NPB{406} (1993) 566.}
\REF\pwr{A. Polyakov and P. Wiegmann, \PLB{141} (1984) 223. \nxl
         N. Reshetikhin and P. Wiegmann, \PLB{189} (1987) 125.\nxl
         G. Japaridze, A. Nersesyan and P. Wiegmann, \NPB{230} (1984) 511.\nxl
         P. Hasenfratz, M. Maggiore and F. Niedermayer,
         \PLB{245} (1990) 522.\nxl
         P. Hasenfratz and F. Niedermayer \PLB{245} (1990) 529.}
\REF\fozo{M. Fowler and X. Zotos, \PR{B24} (1981) 2634 and {\bf B25}
          (1982) 5806.}
\REF\nuev{ Al. B. Zamolodchikov, \PLB{335} (1994) 436 and
Montpellier  preprints 1994.}

\chapter{Introduction}

The computation of thermodynamical functions for integrable models started with
the seminal works of C.N. Yang and C.P. Yang [\yy], of  M. Takahashi  [\tak]
and M. Gaudin [\gaud]. In Refs. [\tak,\gaud] the free energy of the Heisenberg
spin chain is written in terms of the solution of an infinite set of coupled
nonlinear integral equations, derived on the basis of the so--called ``string
hypothesis".

These equation are commonly known as  Thermodynamic Bethe Ansatz
(TBA) equations. Nowadays, the TBA is mostly used as a nonperturbative
tool to investigate Perturbed Conformal Field Theories
and has been considerably generalized (see \eg\ [\albz,\tbam]).
In all cases some some kind of string hypothesis is used.

In ref.[\nos], we proposed a simpler way to solve the
thermodynamics of the XXZ chain and of the sine--Gordon field theory
by means of a single, rigorously derived, nonlinear integral equation.
This method can be applied to a wide class of models solvable by Bethe
Ansatz,
and can provide also the excited--states scaling functions of the
sG field theory [\workinprogress].
Similar methods were used to also determine
some correlation lengths at
finite temperature in the XXZ chain [\klum].

The purpose of the present paper is threefold. First we generalize
the results of ref.[\nos] in the presence of external magnetic field
for the XXZ Heisenberg model.

Second, we show that this approach provides an unified
method to compute both at {\it finite} temperature and infinite size
and at zero temperature and {\it finite} size.
Third, we recast our NLIE in two alternative forms: a) as an infinite
set of algebraic equations of BA type and b) as a Riemann-Hilbert
problem. These forms are appropriate to compute the asymptotic
behaviors around the conformal regime (low $T$ for the magnetic chains and
high $T$ for field theory). In the recursive regime (the opposite to
the conformal one), our NLIE can be easily solved by iteration.

Continuum quantum field theories at finite temperature $T$ are
obtained from periodic boundary conditions (PBC) in the imaginary time with
period $ \b \equiv T^{-1} $. By Wick rotation, this leads to a
periodic spatial box of finite size $\b$. What we achieve here is the
analogous to a Wick rotation on the exact Bethe Ansatz (BA) solution on the
lattice.

Our starting point is the BA diagonalization of a row-to-row
inhomogeneous and twisted transfer matrix
$t(\l,\t_1,\t_2,\ldots,\t_N,\om)$ where there is an inhomogeneity $
\t_n$ attached to each site and a twist $ \om $ in the boundary
conditions ($\om = 0 $ corresponds to PBC).

Setting the  inhomogeneities
$\t_n = 0$ yields the homogeneous six-vertex transfer matrix.

Choosing  the  inhomogeneities alternating $\t_n =  (-1)^{n-1} \t $ yields a
diagonal-to-diagonal transfer matrix $T_N(\t)=t_{2N}(\t,\t,0) $ [\ddvj].
For $\t \to 0,~  T_N(\t)$ gives at order $\t$ the XXZ hamiltonian.
$$
        T_N(\t)\buildrel {\t\to0} \over =
                1-{\t\o{2J\sin\g}}H_{XXZ} + O(\t^2)             \eqn\xxz
$$
and hence,

$$
   e^{-\b H_{XXZ}}= \lim_{N\to\infty} \left[T_L({{2\b J  \sin\g}\o N})\right]^N
$$
This relation permits to write the XXZ partition function
$$
{\Z} \equiv {\rm Tr\,}\left\{ e^{-\b H_{XXZ}} \right\}
$$
in terms of the diagonal-to-diagonal transfer matrix $ T_L(\t) $.
Moreover, using the crossing symmetry of the statistical weights
one can prove that $\Z$ in the infinite volume limit is given just
by the ground state eigenvalue of  $ T_N(\t) $ in the $N \to \infty$ limit.
This alternative way is much simpler than the usual TBA approach, where
one has to sum over an infinite number of states with an arbitrary number
of holes and an arbitrary number of strings of any length.
What we do here is analog to duality in S-matrix theory, where
the sum over an infinite number of resonances in the s-channel
can be replaced by the exchange of one reggeon in the t-channel.

Setting $ \t = -i \Th $ with $\Th $ real, we get the lattice field
theory evolution operators

in the light-cone approach [\ddev ,\ddvj ]

$$
 U_R(\Th)= a(-2i\Th)^{-N}~t_{2N}(-i\Th,-i\Th,0)  ~~,~~
U_L(\Th)= a(2i\Th)^{N}~t_{2N}(+i\Th,-i\Th,0)^{-1}
$$
where $ a(\pm 2i\Th)^{\pm N}$ is a unitarization factor.
That is, for imaginary $\t$, we interpret the lattice as  a discretized
{\bf Minkowski} spacetime [\dev] and hence the operators $U_R$ and $U_L$
describe space and time translations in the light-cone directions of a
discretized Minkowski spacetime. If we call $\d$ the lattice spacing,
$$
    U=e^{-i\d H} \;,\qquad  U_R\; U_L^{-1}=e^{i\d P}    \eqn\evola
$$
define the lattice hamiltonian ($H$) and momentum ($P$) operators.

Moreover, the operator $t_{2N}(\l,-i\Th,0)$ is the generating functional
of local conserved charges (including $H$ and $P$) for $|\Im \l | <
\g/2$ [\ybs].
In the region $-(\pi - \g/2) < \Im \l < -\g/2 $,  $t_{2N}(\l,-i\Th,0)$
generates non-local conserved charges [\ybs].

We diagonalize the transfer matrix $t(\l,\t_1,\t_2,\ldots,\t_N,\om)$
by algebraic Bethe Ansatz (BA). The BA the expresses eigenvectors and
eigenvalues  in terms of the roots of M coupled algebraic equations $(
1 \leq M \leq N)$, the so called Bethe Ansatz equations (BAE).

Using contour integral techniques, we rewrite these BAE as a
non-linear integral equation. This equation may take different forms.
A compact form is
$$
Z(\mu) = \varphi (\mu) + 2\; \Im \intf d\,\mu' \;G(\mu - \mu' -
i\eta) \log\left[1 + e^{iZ(\mu' +i\eta)}  \right]
\eqn\nostra
$$
where  the kernel $ G(\l)$ is explicitly given by eq.(2.24) and
$$
         \varphi (\mu) = \cases{
        2N\arctan \left[{{\sinh(\pi\mu/\g)}\o{\cos(\pi\t/[2\g])}}\right]
        -{{\pi\om}\o{\pi-\g}}   & {\rm finite size vertex model with TBC} \cr
        -{{2\pi \bt}\o{\g\sinh[\pi\mu/\g]}} +  {{i \pi\b h}\o{\pi-\g}}
                                & {\rm XXZ thermodynamics in a
        magnetic field}         \cr
         m\b\, \sinh(\pi\mu/\g) &  {\rm sine-Gordon field theory} \cr}
                                                          \eqn\diffi
$$
$Z(\mu)$ is the unknown
in eq.\nostra . The real parameter $ \eta $ can be freely chosen in the
interval $ 0 < \eta < \g / 2 $ .

We have unified three different problems
(finite size vertex model with TBC,  the XXZ thermodynamics in a
magnetic field and the sine-Gordon field theory in a finite volume)
into the single NLIE \nostra -\diffi .

Once  we have the solution $Z(\l)$ of eq.\nostra , we can compute the
transfer matrix eigenvalues by quadratures. We consider three relevant
applications

\item{a)} The finite size (homogeneous) six-vertex model with twisted
boundary conditions (TBC).

\item{b)} The (infinite) XXZ chain at temperature $T= \b^{-1}$ in an external
magnetic
field $h$.

\item{c)} The massive Thirring-sine Gordon  model (mT-sG) on a
periodic box of size $\b$.

In the first case, we find for the six vertex partition function on a
double periodic square lattice ($T$ x $2N$ )
$$
    \lim_{T\to\infty} T^{-1}\log \Z_{6V} = \log\tau_{max}
              = -2N f_0(\l,\om)+ L_{2N}(\l,\om)          \eqn\lnpar
$$
For large $N$, we obtain that $L_N$  follows the conformal behaviour
for  TBC with unit central charge [see sec. 7.1]. Namely,
$$
L_{2N}\buildrel {N\to \infty} \over  =
{{\pi}\o{6N}} ~ \tan\left({{\pi \l}\o {\g}}\right)~
\left[1 - {{6\om^2}\o {\pi^2 (1 - \g/\pi)}} \right]
$$
In the case b) , we relate the XXZ free energy at temperature $T$
and on an external magnetic field $2h$ in the $z$-direction with the
maximal eigenvalue of the alternating transfer matrix
$t_{2N}(\pm\t,\t,-i\b h)$ with TBC. Notice that the twist, $-i\b h$, is
purely imaginary for real magnetic fields $h$. This new result allows
us to express the XXZ free energy $f(\b,h)$ in terms of the solution
$Z(\l)$ of our non-linear integral equation (NLIE) \nostra .
$$
        f(\b,h)=h+2J\cos\g +{{\sin\g}\o\b}\oint_\Gamma {{d\l}\o{2\pi i}}\,
        {{\log\left[ 1 + e^{-iZ(\l)}\right]} \o
        {\sinh\l\,\sinh(\l-i\g)}}                       \eqn\elibre
$$

Finally, we use the light-cone approach [\ddev, \ddvj , \dev , \ybs] to
deal with c). We write the energy in the mT-sG model on a ring of
length $\b$ in terms of the alternating six vertex transfer matrix
[$t(\mp i\Th,-i\Th,0)$]
eigenvalues. The basic relation is the the formula for the unit
evolution operator for  lattice  field models,
$$
        U= e^{-i\d H} = \left[{{a(2i\Th)}\o{a(-2i\Th)}}\right]^N  ~
                t(-i\Th,-i\Th,0)~ t(i\Th,-i\Th,0)^{-1}  \eqn\basic
$$
where $\d$ is the  lattice spacing, $\Th$ plays the r\^ole of UV
cutoff and $a(\pm2i\Th)$ are unitarizing c-number factors.
In the continuum limit, $\d \to 0,~N\to \infty $ with $ \b \equiv N \d
$ fixed. At the same time $\Th$ tends to infinite as
$$
        \Th \simeq {\g\o\pi}\log{{4N}\o{m\b}}                   \eqn\teta
$$
We find for the scaling function
$$\eqalign{
    E(\b) &\equiv \lim_{N\to\infty}\left[ E_N-E_c \right] \cr
        &=-{m\o\g} \,\Im \!\intf {{d\l}\o \pi}\,
\sinh\left[\pi(\l+i\eta)/\g \right] \log\left[
            1+e^{iZ(\l+i\eta)}\right]           \cr}            \eqn\enbe
$$
where $Z(\l)$ satisfies [cfr. eq.\nostra ]
$$
    Z(\l)=m\b\sinh(\pi\l/\g)+2~\Im \intf d\mu\,G(\l-\mu-i\eta)
        \log\left[ 1+e^{iZ(\mu+i\eta)}\right]                   \eqn\netba
$$
provided we choose $0<\eta<\frac12 {\rm min}(\g,\pi-\g)$.

Shifting $\l$ by $i\g/2$, eq.\netba\ takes the form
$$
        \e_f  = m\b\cosh\t - G_0*L_f + G_1*L_\fb
\eqn\oursi
$$
where
$$\eqalign{
        & \e_f(\t) = -iZ(\frac\g\pi \t+i\g/2)
                \;,\qquad \e_\fb =\overline{\e_f}       \cr
        & L_f \equiv \log(1+e^{\e_f}) \;,\qquad
                L_\fb \equiv \log(1+e^{\e_\fb}) =  \overline{L_f} \cr
        & G_0(\t)=\frac\g\pi G(\frac\g\pi \t)
                \;,\quad G_1(\t)=G_0(\t+i\pi-i0)   \cr}         \eqn\idef
$$
 The kernels  $G_0(\t)$ and   $G_1(\t)$ are just the $\t$ derivative of the
fermion-fermion and fermion-antifermion physical phase-shifts.
This  transparent physical interpretation of  eq.\oursi\
suggests the form that it should take in other field theory models.

Now we face the problem of solving NLIE like \nostra\ or \netba\
in order to get physical results. It must be noticed that we find just
{\bf one} NLIE for finite temperature and finite size situations.
This situation contrasts with the standard TBA
which for nonrational values of $\g/\pi$ involves an infinite number of
nonlinear integral equations for the infinitely many different types of
``magnons" describing the energy degeneracies of the physical fermions and
antifermions [\fozo]. Therefore, {\bf our equation effectively provides
 a resummation of the magnon degrees of freedom}.

There are essentially two regimes in our NLIE. We call them `recursive'
and `conformal'. In the  `recursive' regime the NLIE
can be simply solved by iterating the inhomogeneity.
This happens for small $\b$ in the XXZ chain and for
large   $\b$ in field theory.

The `conformal' regime is the opposite to the  `recursive' regime. In
it, the calculations are more subtle. The dominant behaviour (large
$\b$ for the XXZ chain and small $\b$ in field theory), follows in
closed form with the help of the Lemma in sec. 7. This lemma avoids
the detailed resolution of the NLIE for the dominant contributions.
Higher corrections follow using the Riemann-Hilbert form of our NLIE
(sec. 7.4).  We  find that the energy in the sG-mT model has
a series of terms with
$(m\b)^{-1}$ times integer powers of $(m\b)^{4\g/\pi}$ plus the central charge
term  proportional to $(m\b)^{-1}$,
in agreement with the predictions of Perturbed Conformal Field Theory.

Since the inhomogeneity in the NLIE for field theory is $m\b \cosh\t$,
it is useful to define for small $m\b$ kink solutions, $\e_k(\t)$ of the NLIE
with inhomogeneity $e^{\pm\t}$ [see eq.(7.19)]. We obtain in this way for
the energy $E(\b)$ at high temperatures
$$
        E\buildrel {\b\to 0} \over = -{{\pi}\o{6\b}}
        - {{m^2\b}\o 4}\,{\rm cot}\left[\,\pi^2/2\g \right]
         -{m\o{2\pi}}\intf d\t\,\cosh\t
        \left[\ell(\t)+\bar{\ell(\t)}\right]    \eqn\enkink
$$
The first term is the universal conformal Casimir energy, from which we read
the correct central charge $c=1$ of the mT-sG model.
The second term, linear in $\b$, exactly coincides with minus the scaling
bulk free energy [see eq.(5.7)].
The last integral in eq.\enkink\  represents the resummation of the
perturbed conformal field theory around the massless Thirring model.
The fact that it must contain non-integral
powers of $r$, causing non-analyticity at $r=0$,
can be established directly from the original equation (7.18),
since this cannot be expanded in a Taylor series of $r$.

In the  `recursive' regime, we derive in sec. 6.1, the high
temperature expansion for the XXZ chain in the external field $h$.
One {\bf only needs the Cauchy theorem} to evaluate the high $T$ expansion
coefficients. The first five terms are explicitly computed in
sec. 6.1:
$$\eqalign{
       f(\b)=&-\b^{-1}\ln2 +J\cos\g-\b \left[
         J^2(1+\frac12{\rm cos}^2\,\g)+\frac12 h^2 \right]
                +\b^2\cos\g (J^3+h^2) \cr
        &- \b^3 [(\frac14 +\frac16 \cos^2\g -\frac1{12}\cos^4\g)J^4 -
                J^2 h^2 -\frac1{12} h^4 ] +O(\b^4)  \cr}
$$
which indeed agrees with the high $T$ expansion.

In sec. 6.2  we obtain the ground state energy for
the mT-sG model for small $T$. It is non-analytic at $T=0$.
We find up to contributions $ O(e^{-3m\b})$
$$
E(\b) \buildrel{\b\to\infty}\over= -{{2m}\o\pi}K_1(m\b)+
 e^{-2m\b}       \sqrt{{m\o{\pi \b}}} \left[ {{1+\sqrt 2}\o 2}
+\sqrt{{\pi}\o{m\b}} \; K_{\g} + O({1\o \b}) \right]  + O(e^{-3m\b})
\eqn\sgbt
$$
where $K_1(x)$ stands for a modified Bessell function and the constant
$K_{\g}$ is defined by the integral (6.17).
The first term in eq.\sgbt\ describes
free massive bosons, the subsequent terms take into account the
interactions.

In secs 4 and 5, we show that our NLIE \nostra\ and \netba\ are
equivalent to an infinite set of algebraic equations of BA type. These
equations differ from the usual BA equations [\fata , \dev ] in two
essential aspects: i) they are infinite in number, ii) the roots are
discretely spaced although they describe models with an infinite
number of sites. The root spacing tends to zero only in the conformal limit.

We would like to remark that, contrary to the traditional
thermodynamic  Bethe Ansatz
[\tak,\gaud], our approach does not rely on the string hypothesis on the
structure of the solutions of the BA equations characteristic of the
Heisenberg chain. This makes our approach definitely simpler. Most notably,
the whole construction of the thermodynamics no longer depends on whether
$\g/\pi$ is a rational or not, unlike Takahashi approach [\tak]. This
applys equally well to the problem of the ground state scaling function of the
sine--Gordon field theory, since the standard TBA approach requires the
string hypothesis.

Generalizations of the present TBA approach to higher spin chains as
well as to magnetic chains and quantum field theories associated to Lie
algebras other than $A_1$ (nested BA solutions) are relatively
straightforward provided crossing symmetry holds.

 Our NLIE has been recently used with success in ref.[\nuev].

\chapter{Nonlinear Integral Equation }

Let us consider
a horizontal line formed by N sites, and  associate to each site
$n=1,2,\ldots, N$ a triplet of (vertical) Pauli matrices
$(\s^x_n,\,\s^y_n,\,\s^z_n)$, with standard commutation rules
$$
\left[\s^\a_m,\,\s^\b_n\right] =2i\,\delta_{mn}\,\e^{\a\b\g}\,\s^\g_n
$$
As customary, we also introduce the auxiliary, or horizontal,
two--dimensional space, labeled by the index $n=0$, on which the
matrices $(\s^x_0,\,\s^y_0,\,\s^z_0)$ act. For any complex value
$\l$ of spectral parameter and an arbitrary set of
inhomogeneities $\t_1,\t_2,\ldots,\t_N$, the local vertices
$L_n$ are then written as
$$
        L_n =R_{0\,n}(\l+\t_n)\, P_{0\,n}
\eqn\localv
$$
in terms of the spin $1/2$ $R-$matrices of the six vertex model,
$$
        R_{k\,n}(\t)=  {{a+c}\o2}+{{a-c}\o2}\s^z_k\s^z_n +
                {b\o2}\left( \s^x_k\s^x_n + \s^y_k\s^y_n\right) \eqn\rma
$$
and of the exchange operators,
$$
        P_{k\,n} = \frac12 \left(1+ {\vec\s}_k \cdot {\vec\s}_n \right)
$$
The dependence on $\t$
of $R_{k\,n}(\t)$ is contained in the the Boltzmann weights $a,b,c$, for
which we take the standard trigonometric parametrization
$$
        a =a(\t) \equiv \sin(\g-\t)  \;,\quad
        b =b(\t) \equiv \sin \t \;,\quad c=\sin\g               \eqn\param
$$
where $\g$ is the anisotropy parameter. Throughout this paper, unless
explicitly stated otherwise, we shall confine our attention to $\g$ real
and in the interval $(0,\pi/2)$ (by periodicity, the complete relevant
interval is $0\le \g \le \pi$).

The inhomogeneous and twisted six--vertex transfer matrix associated
to this N--sites line reads
$$
        t(\l,\t_1,\t_2,\ldots,\t_N,\om) = e^{i\om\s^z_0}~
                {\rm tr}_0\, \left[ L_1 L_2 \ldots L_N  \right] \eqn\alter
$$
The twist angle $\om$ defines the relation between spin operators
after a translation by N sites:
$$
    \s^x_{n+N} \, \pm \, i \s^y_{n+N} \, = e^{\pm i\om}~[ \s^x_n \, \pm
    i\s^y_{n} ] \; , \qquad \s^z_{n+N} = \s^z_n      \eqn\twis
$$
Notice also that $R_{k\,n}$ is $U(1)-$invariant,
$$
        e^{i\phi(\s_k^z+\s_n^z)}~ R_{k\,n}(\t)=
        R_{k\,n}(\t) ~ e^{i\phi(\s_k^z+\s_n^z)}                 \eqn\uone
$$
so that the transfer matrix commutes with the $z-$projection of the
total spin $S^z=\frac12 \sum_1^N \s^z_n$.

The diagonalization of $t(\l,\t_1,\t_2,\ldots,\t_N,\om)$ is achieved via
Quantum Inverse Scattering
Method (also called Algebraic Bethe Ansatz) (ref. [\fata ,\dev ]):
the eigenstates with $S^z=N/2-M$, $M=0,1,2,\ldots,[N/2]$, are
labelled by the $M$ distinct roots $\{\l_1,\l_2,\ldots,\l_M\}$ of the Bethe
Ansatz Equations (BAE)
$$
        \prod_{n=1}^N
        {{\sinh(\l_j+i\t_n+i\g/2)} \o {\sinh(\l_j+i\t_n-i\g/2)}}
                        =- e^{-2i\om} \prod_{r=1}^M
        {{\sinh(\l_j-\l_r+i\g)} \o {\sinh(\l_j-\l_r-i\g)}}      \eqn\bae
$$
while the corresponding eigenvalues $\tau(\l,\t_1,\t_2,\ldots,\t_N,\om)$ read
$$\eqalign{
        \tau(\l,\t_1,\t_2,\ldots,\t_N,\om)      &=
        e^{i\om} \prod_{n=1}^N a(\l+\t_n)  \;\; \prod_{r=1}^M
        {{\sinh(i\g/2-\l_j+i\l)} \o {\sinh(i\g/2+\l_j-i\l)}}    \cr
 &+     e^{-i\om} \prod_{n=1}^N b(\l+\t_n)  \;\; \prod_{r=1}^M
        {{\sinh(3i\g/2+\l_j-i\l)} \o {\sinh(-i\g/2-\l_j+i\l)}}\cr}  \eqn\eee
$$
We shall primarily be interested in one particular solution with $S^z=0$
of the BAE, the so--called antiferromagnetic ground state (a.g.s.) solution,
which will be characterized shortly. Since the request $S^z=0$ forces $N$ to be
even, we replace $N$ by $2N$ throughout. Moreover, for our applications, we
shall need to consider only a special type of inhomogeneity structure, the
alternating one
$$
                \t_n = (-1)^{n-1} \t/2                          \eqn\altt
$$
where $\t$ will either be real, with $0 \le \t <\g/2$ , or purely imaginary.
Correspondingly, the object of our interest is the (twisted) alternating
transfer matrix
$$
        t_{2N}(\l,\t,\om) =t(\l,\t_1=\t/2,\t_2=-\t/2,\ldots,\t_{2N}=-\t/2,\om)
                \eqn\alttra
$$
As is well known, this object is a {\bf diagonal to diagonal} transfer
matrix with $N$ links in the horizontal direction [\ddvj].
Suppose now that $N$ distinct roots
$\{\l_1,\l_2,\ldots,\l_N\}$ are given; with them we construct the
so--called {\it counting function} [\dev]
$$
        Z_N(\l)= N\left[\phi(\l+i\t/2,\g/2)+\phi(\l-i\t/2,\g/2) \right] -
        \sum_{k=1}^N \phi(\l-\l_k,\g) -2\om                     \eqn\count
$$
where$$
        \phi(\l,x)\equiv i\log{{\sinh(ix+\l)}\o{\sinh(ix-\l)}} \; \qquad
        , \qquad        (\phi(0,x)=0)                   \eqn\defi
$$
has the cut structure chosen so that it is analytic in the strip
$|\Im\l| \le x$. The a.g.s. solution is now specified by the property
$$
        Z_N(\l_j)=(-N+2j-1)\pi \;,\qquad j=1,2,\ldots,N         \eqn\prop
$$
and enjoys the symmetry (compare with the BAE, eq.\bae)
$$
        \l_j(\om)=\overline{\l_j({\bar\om})}=-\l_{N-j+1}(-\om)  \eqn\symm
$$
The existence of this solution is most simply established
by numerical calculations, for $N$ up to the thousands. The uniqueness
will become apparent in the sequel. For $\om$ real, all $\l_j$ are also
real and the a.g.s. can be characterized as the unique solution of the BAE with
$N$ real roots. If $\Im\om>0$ ($<0$)
then $\Im\l_j>0$ ($<0$) and the a.g.s. goes continuously, by construction, into
the real solution as $\Im\om\to 0$. Notice that the symmetry \symm\ implies for
the counting function itself
$$
        Z_N(\l,\om)=\overline{Z_N({\bar\l},{\bar\om})}
                   = -Z_N(-\l,-\om)                     \eqn\refl
$$
which will play an important r\^ole in the applications. Notice also that all
these last statements really hold true for $\g$ in the extended interval
$(0,\pi)$.

To gain some understanding on the distribution of roots in this a.g.s. solution
one can try to approximate it in the limit of $N$ large and fixed
(\ie\  $ N-$independent) $\t$ and $\om$. The idea is the traditional one that
the BA root distribution becomes the continuous density $\s_c(\l)$ (normalized
to 1) in the limit $N\to\infty$.  Then the sum over the roots in the
definition of $Z_N(\l)$, eq.\count, can be approximated by $N$ times the
integral over the density $\s_c(\l)$. At the same time the property
\prop\ implies that
$$
  \lim_{N\to\infty} {1\o N}{d\o{d\l}} Z_N(\l) = 2\pi\s_c(\l)    \eqn\limcount
$$
so that one finds the following linear integral equation for $\s_c(\l)$
$$
        2\pi\s_c(\l)=\phip(\l+i\t/2,\g/2)+\phip(\l-i\t/2,\g/2) -
        \intf d\mu\,\phip(\l-\mu,\g)~\s_c(\mu)                  \eqn\linear
$$
where $\phi^\prime(\l)=d\phi(\l)/d\l$.
Fourier transforming leads to the explicit solution
$$
        \s_c(\l) = z_c^\prime(\l)   \; ,\qquad
        z_c(\l)= {1\o\pi} \tan^{-1} \left[
        {{\sinh(\pi\l/\g)}\o{\cos(\pi\t/[2\g])}}\right]         \eqn\zetac
$$
yielding to leading order $Z_N(\l)\simeq  2\pi N z_c(\l)$. Notice that no
dependence on the twist parameter $\om$ is left over in this continuum
approximation. One can attempt to recover $\om$ by adding to $2\pi N
z_c(\l)$
the constant function $f_0(\l)=-2\om$, as suggested by the definition itself of
$Z_N(\l)$, eq.\count. However, this continuum approximation need not control
the $O(1)$ terms. In fact, to solve \linear\ one has to invert the convolution
$(1+K)*\s_c$, where $K$ is defined as
$$
        (K*f)(\l) \equiv \intf{dx\o{2\pi}} \phip(\l-x,\g)f(x)   \eqn\convo
$$
This implies that the correct constant to add is $(1+K)^{-1}*f_0$, that is
$$
         2\pi N z_c(\l) -{{\pi\om}\o{\pi-\g}}   \simeq Z_N(\l)
\eqn\appr
$$
Therefore in the continuum approximation the roots are given by
$$
        \l_j \simeq {\g\o\pi} \sinh^{-1} \left[\cos{{\pi\t}\o{\g}} \,
        \cot {{\pi}\o{2N}} (2j-1 +{{\om}\o{\pi-\g}})\right]     \eqn\apprlam
$$
Although imprecise, this approximation is useful, for
numerical tests show that the actual values of $\l_j$ are very close for
arbitrary values of $\t$ and $\om$. Thus we see that the roots lay on a
smooth curve entirely contained in some horizontal strip $\D$ of the
the upper (lower) complex plane for $\Im\om>0$ ($\Im\om<0$).
In turn, $\D$ is included inside the strip $\Im\l<\g/2$.
Moreover, from a host of numerical checks
as well as from arguments based on the unicity of the a.g.s.
configuration, we expect to find no other point, in the strip
$|\Im \l|< \g/2$, where $e^{iZ_N}$ takes the value $(-1)^{N+1}$,
besides the roots
$\l_j$. We shall now look for the modifications to the linear eq.\linear,
which are necessary to go beyond the continuum approximation.

For sake of simplicity, we shall consider the case $N$ even (after all
$N\to\infty$ in physical applications) and $\Im\om>0$ (recall the property
\refl). By construction, the function $\exp[iZ_N(\l)]$ is analytic
in the strip $(-\g+\t)/2< \Im \l< \g/2$ and the sum in the last term of
eq.\count\ can be written as a contour integral
$$
        \sum_{j=1}^N \phi(\l-\l_j,\g) = \oint_\Gamma {{d\mu}\o{2\pi i}}\,
        \phi(\l-\mu,\g)  {d\o{d\mu}} \log[1 + e^{iZ_N(\mu)}]    \eqn\sumi
$$
where the closed contour $\Gamma$ encircles counterclockwise all the roots
$\l_j$ (recall that, by construction, $\exp[iZ_N(\l_j)]=- 1$).
We can take $\Gamma$ to run
straight from left to right at $\Im z=-\eta_-$, $0 <\eta_- <(\g-\Re\t)/2$ and
come back straight at $\Im z=\eta_+$, ${\rm max}(\Im\l_j) <\eta_+ <\g/2$
(of course the exact values of $\eta_-$ and $\eta_+$  do not matter because of
the analyticity and the behaviour at real infinity of $e^{iZ_N}$).

Inserting eq.\sumi\ in eq.\count, using $1+e^{iZ_N}=e^{iZ_N}(1+e^{-iZ_N})$
and integrating by parts yields
$$\eqalign{
        Z_N(\l) &+\intf{dx\o{2\pi}}\, \phip(\l-x,\g) Z_N(x) =
        N\left[\phi(\l+i\t,\g/2)+\phi(\l-i\t,\g/2) \right] -2 \om       \cr
                & -i\intf {dx\o{2\pi}}\, \phip(\l-x-i\eta_+,\g)
        \log \left[ 1+e^{iZ_N(x+i\eta_+)} \right]               \cr
                & +i\intf {dx\o{2\pi}}\, \phip(\l-x+i\eta_-,\g)
        \log \left[ 1+e^{-iZ_N(x-i\eta_-)} \right]   \cr}       \eqn\segunda
$$
Notice that the function $\exp[-iZ_N(\l)]$ is analytic
in the strip $(\g-\Re\t)/2> \Im \l>-\g/2$, so that $\eta_-$ may be chosen
to vary up to $\g/2$. We next introduce the convolution operation

$$
        G = K*(1+K)^{-1} \;,\quad
   G(\l) =\intf{{dk}\o{4\pi}}{{\sinh(\pi/2-\g)k}\o
              {\sinh(\pi-\g)k/2 \,\cosh(\g k/2)}}~e^{ik\l}      \eqn\kernel
$$
so that eq.\segunda\ can be rewritten
$$\eqalign{
        Z_N(\l) &= 2\pi N z_c(\l) - {{\pi\om}\o{\pi-\g}}
                -i\intf dx\, G(\l-x-i\eta_+)
                \log \left[1+e^{iZ_N(x+i\eta_+)} \right]  \cr
                &+ i\intf dx\, G(\l-x+i\eta_-) \log
                \left[1+e^{-iZ_N(x-i\eta_-)} \right]      \cr}    \eqn\terza
$$
This expression is exact and exhibits (although only implicitly) all the
corrections to the continuum approximation \appr. If $\Im\om=0$, so that the
roots $\l_j$ are real, the property $Z_N({\bar\l})=\overline{Z_N(\l)}$ allows
to interpret the identity \terza\ as a nonlinear integral equation for the
counting function itself. In fact, setting $\eta_+=\eta_-\equiv \eta$,
$Z_N(\l+i\eta) =i\e(\l)$, $G(\l+iy)=G_y(\l)$  and
$$
 g(\l) =-2\pi i N z_c(\l+i\eta) + {{i \om}\o {1 -\g/\pi}}
\eqn\gedel
$$
we obtain, in compact notation
$$
    \e = g - G*\log(1+e^{-\e}) + G_{2\eta}*\log(1+e^{-{\bar\e}})  \eqn\compact
$$
We recall that $\eta$ can be chosen freely in the interval $0<\eta<\g/2$,
since the original assumption $0<\g\le\pi/2$ implies that $G(\l)$ is analytic
in the strip $|\Im\l|<\g$. Eq. \compact\ may nontheless be extended to the
other region $\pi/2<\g<\pi$, where $G(\l)$ is analytic for  $|\Im\l|<\pi-\g$,
simply by requiring that $0<\eta<\frac12{\rm min}(\g,\pi-\g)$.
For $\eta\to\g/2$, eq.\compact\ agrees with the nonlinear integral equation
derived in ref. [\aus] by different methods.

To convert eq.\terza\ into a constructive integral equation also when
$\Im\om\ne 0$ we can exploit the symmetry \refl, namely
$Z_N({\bar\l},{\bar\om})=\overline{Z_N(\l,\om)}$ and combine the two cases
$\Im\om>0$ and $\Im\om<0$. In fact, the change $\om\to {\bar\om}$ simply
amounts to change the restrictions on $\eta_\pm$, which now read $0 <\eta_+
<\g/2$ and $-{\rm max}(\Im\l_j) <\eta_- <\g/2$. It is again convenient to fix
$\eta_+=\eta_-=\eta$, although the lower bound for $\eta$  actually depends on
the roots and introduces an extra nonlinearity
in the problem. The choice $\eta=\g/2$, however, (recall that we assumed
$0<\g<\pi/2$) would eliminate this subtlety. Then we set
$$\eqalign{
                Z_N(\l+ i\eta;\Re\om \pm i \Im\om) &=i \e_\pm(\l) \cr
                2\pi Nz_c(\l+i\eta)- {\pi\o{\pi-\g}}(\Re\om \pm i \Im\om)
                &=i g_\pm(\l) \cr}                      \eqn\defeps
$$
and arrive at the system of two coupled nonlinear integral equations
$$
        \e_\pm = g_\pm - G*\log(1+e^{-\e_\pm}) +
                G_{2\eta}*\log(1+e^{-{\bar\e}_\mp})             \eqn\compactpm
$$
Let us recall that for $\Im\om=0$ we
can choose $\eta\to\frac12{\rm min}(\g,\pi-\g)$, while for nonzero $\Im\om$
and $\g<\pi/2$ we can fix $\eta\to\g/2$. This choices are assumed from now on,
unless explicitly stated otherwise.

\chapter{Finite size corrections}

The nonlinear integral equations \compactpm\ are {\bf exact} for any
N. In the large N limit they allow to find the finite size corrections
to the eigenvalue \eee\ of the transfer matrix for the a.g.s.
configuration. For simplicity we shall consider here the
homogeneous case (\ie\  with $\t=0$) which is relevant in the
calculation of the partition function  $\Z_{6V}$ of the standard six--vertex
model on a very long torus. Indeed on a doubly periodic square lattice
with $T$ sites in the vertical direction and $2N$ sites in the
horizontal direction $Z_{6V}$ reads
$$
     \Z_{6V}= {\rm Tr}\left[ t_{2N}(\l,0,\om) \right]^T       \eqn\parti
$$
Actually, this periodic lattice has a vertical seam across which the
vertical statistical variables $\s=\pm$ acquire the factor
$e^{\pm i\om}$, as evident from eq.\twis. To simplify the discussion,
in this section we shall take $\om$ to be real.

In the cylinder limit $T \to \infty$ at fixed $N$, the largest
eigenvalue $\tau_{max}$ of the transfer matrix dominates and we have
$$
    \lim_{T\to\infty} T^{-1}\log \Z_{6V} = \log\tau_{max}
              = -2N f_0(\l,\om)+ L_{2N}(\l,\om)          \eqn\logparti
$$
where $f_0$ is the free energy per site in the infinite $N$ limit,
and $L_{2N}$ represents the
finite size corrections. With our original choice of a real anisotropy
$\g$, the six vertex model is in the
critical regime, so that we
expect $L_{2N}$ to be governed for large $N$ by conformal invariance.

It is well known that the $\tau_{max}$
exactly corresponds to the a.g.s. configuration analyzed in the
previous section, so our problem is now to relate $\tau_{max}$ to
the solution of the integral equation \compact. To this end we  first
notice that, provided $0\le \Re\l<\g/2$, the second
term in the expression \eee\ for the a.g.s.
eigenvalue is exponentially small relative to the first one (for an
explicit verification see eq.(6.29) of ref. [\ybs ]), and may therefore
be dropped. Thus we can write
$$
    \log\tau_{max}= i\om + 2N \log a(\l) + E_N          \eqn\taumax
$$
where
$$
       E_N =  -i\sum_{j=1}^N \phi(\l_j +i\l,\g/2)        \eqn\enerI
$$
The sum over the roots can be transformed into integrals by the
same procedure used in sect. 2 for the counting function, with the
result (we must choose  here $\eta_-<\g/2-\Re\l$)
$$\eqalign{
  E_N= &-i\intf {{d\mu}\o{2\pi}}\, \phi^\prime(\mu+i\l,\g/2)\;Z_N(\l)
        -\intf {{d\mu}\o{2\pi}}\, \phi^\prime(\mu+i\l+i\eta_+,\g/2)
\log \! \left[1+e^{iZ_N(\mu+i\eta_+)} \right] \cr
        &+\intf {{d\mu}\o{2\pi}}\, \phi^\prime(\mu+i\l-i\eta_-,\g/2)
\log \! \left[1+e^{-iZ_N(\mu-i\eta_-)} \right] }\eqn\enerII
$$
Eq. \terza\ may now be used to eliminate the term linear in $Z_N(\l)$,
yielding (we choose $\eta_+=\eta_-\equiv \eta$ as usual)
$$\eqalign{
   E_N = E_c &-\intf d\mu\, \left\{ \s_c(\mu+i\l+i\eta)
        \log\!\left[1+e^{iZ_N(\mu+i\eta)} \right] \right.   \cr
         &- \left.  \s_c(\mu+\l-i\eta)
        \log\!\left[1+e^{-iZ_N(\mu-i\eta)} \right] \right\}\cr}\eqn\enerIII
$$
where $E_c$ is the continuum  approximation
$$\eqalign{
    E_c= & -iN\intf d\mu\, \phi(\mu+i\l,\g/2)\;\s_c(\mu)     \cr
       = & \; 2N \;  \intf {{dk} \o k } \;
{{\sinh[(\pi - \g)k] \sinh (2 \l k)} \o {\sinh(\pi k ) \cosh(k \g)}}~
,~|\l| < \g/2 } \eqn\econtI
$$
and (recall eq.\zetac\  and that $\t=0$ here)
$$
      \s_c(\mu) = {1\o{\g\cosh(\pi\mu/\g)}}=
                   {1\o{2\pi}}[(1-G)\ast \phi^\prime](\mu)  \eqn\sigmac
$$
Comparing eqs. \logparti\ and \enerIII\ we obtain
$$
f_0 = - {1 \o {2N}}\;E_c + \log a(\l)
$$
[notice that $ E_c/N $ is $N$-independent] and the finite size
corrections expressed in terms of  the solution
of the nonlinear integral equation \compact\ :
$$
        L_N(\l,\om)=  i\om -{1 \o {\g}}\intf d\mu\, \left\{
        {{\log[1+e^{-\e(\mu)}]} \o {\cosh[\pi(\mu+i\l+i\eta)/\g]}} -
  {{\log[1+e^{-\overline{\e(\mu)}}]} \o {\cosh[\pi(\mu+i\l-i\eta)/\g]}}
                     \right\}                             \eqn\fscorr
$$
In section 7.1 we compute the dominant large $N $ behaviour of
$L_N(\l,\om) $ which yields the central charge.

\chapter{Thermodynamics of the XXZ spin chain}

The Hamiltonian of the periodic XXZ spin chain with $2L$ sites reads
$$
      H_{XXZ}=-J\sum_{n=1}^{2L} \left[
      \s^x_n\s^x_{n+1} + \s^y_n\s^y_{n+1} -\cos\g\,(\s^z_n\s^z_{n+1}+1)
        \right]                                                   \eqn\hxxz
$$
and is very simply related to the transfer matrix $T_L(\t)$ of the symmetric
six--vertex model on a diagonal lattice with $2L$ links in the "space"
direction. In fact $T_L(\t)$, with a suitable normalization, can be written
$$
        T_L(\t)=  c^{-2L} R_{2\,3} R_{4\,5} \ldots R_{2L\;1}
                R_{1\,2} R_{3\,4} \ldots R_{2L-1\;2L}           \eqn\tran
$$
with the $R-$matrices $R_{k\,n}$ given in eq.\rma, and
for small $\t$ one finds
$$
        T_L(\t)\buildrel {\t\to0} \over =
                1-{\t\o{2J\sin\g}}H_{XXZ} + O(\t^2)             \eqn\ttoh
$$
Notice that the higher order local operators which follow from the
$\t$ expansion do not commute among themselves nor with $H_{XXZ}$,
since diagonal--to--diagonal transfer matrices do not commute for different
values of $\t$. This is the reason why it is preferable not to call $\t$
spectral parameter in this context.

We now observe that eq.\ttoh\ allows to write
an euclidean path--integral relation between the classical partition function
of the six--vertex model and the quantum partition function of the XXZ
model [\suz,\japs, \taka].
Let us consider the XXZ free energy per site, $f(\b,h)$, in the presence of
an external constant field $2h$ coupled to the conserved $z-$projection of the
total spin $S^z=\frac12 \sum_n \s^z_n$,
$$
      f(\b,h)=-{1\o\b}\lim_{L\to\infty}{1\o{2L}}\log
   \left[{\rm Tr\,}\left\{ e^{-\b(H_{XXZ}-2hS^z)} \right\}\right]    \eqn\free
$$
Setting $\bt=\b J\sin\g$, from eq.\ttoh\ we then read
$$
   e^{-\b H_{XXZ}}= \lim_{N\to\infty} \left[T_L(2\bt/N)\right]^N
$$
Hence we can write
$$
     f(\b,h)=-{1\o\b}\lim_{L\to\infty}{1\o{2L}}\lim_{N\to\infty}
           \log \Z_{LN}(2\bt/N,h)                                   \eqn\ztof
$$
where
$$
\Z_{LN}(\t,h) \equiv {\rm Tr\,}\left\{ e^{2\b hS^z} [T_L(\t)]^N \right\}
\eqn\part
$$
is the six--vertex partition function on a square periodic diagonal lattice
with $2NL$ sites, with the insertion of an horizontal seam along which
each link variable $\s=\pm$ is
multiplied by $e^{\pm\b h}$. Notice that any two neighboring elements of
this seam can be freely moved through a common vertex thanks to the
$U(1)$ invariance of the $R-$matrices, eq.\uone.
The two limits in eq.\part\ cannot be interchanged since the degeneracy
of $T_L(0)=1$, that is $2^{2L}$, is strongly $L-$dependent. However, the
numerical value of $\Z_{LN}(\t,h)$ does not change under
a rotation by $\pi/2$ of the entire lattice, nor under the flipping from
$\pm$ to $\mp$ of the value of all link variables on, say, the left--oriented
diagonals. On the other hand these two operations combined are equivalent to
the substitution $\t\to\g-\t$ in each local $R-$matrix $R_{j\,k}$
(this is called ``crossing symmetry'', and $\g$ is then identified as crossing
point, see $e.g.$ ref.[\baxt]), so that we can write
$$
              \Z_{LN}(\t,h)=\Z_{NL}(\g-\t)_h =
         {\rm Tr\,}[T_N(\g-\t)_h ^L]                    \eqn\cross
$$
where $T_N(\t)_h$ is the twisted diagonal--to--diagonal transfer matrix
$$
  T_N(\t)_h= c^{-2N} e^{ \b h\s^z_1}  R_{2\,3} R_{4\,5} \ldots R_{2N\,1}
              e^{-\b h\s^z_1} R_{1\,2} R_{3\,4} \ldots R_{2N-1\,2N}
\eqn\twtran
$$
Combining eqs. \ztof\ and \cross, we now obtain
$$
     f(\b,h)=-{1\o\b}\lim_{L\to\infty}{1\o{2L}}\lim_{N\to\infty}
           \log \Z_{NL}(\g-2\bt/N)_h                           \eqn\crossed
$$
It is well known [\ddvj] that $T_N(\t)$ is simply related to the
row--to--row (untwisted) alternating transfer matrix $t(\l,\t,\om=0)$
introduced in section 2
$$
  T_N(\t) = [a(\t)a(-\t)/c^2]^N t(\t/2,\t/2,0)
        \; t(-\t/2,\t/2,0)^{-1} \eqn\basic
$$
The presence of the external field can be related to {\bf twisted boundary
conditions}  [eq.\twis ] in absence of magnetic field.
We consider a {\it purely imaginary twist} $\om = -i\b h$, and find
quite  simply
$$
       T_N(\t)_h =      [a(\t)a(-\t)/c^2]^N
                t(\t/2,\t/2,-i\b h)\; t(-\t/2,\t/2,-i\b h)^{-1}  \eqn\basich
$$
Therefore the eigenvalues of $T_N(\t)_h$ can be read off the general result
\eee\ of the Algebraic BA, with the substitutions
$$
  \l=\t/2 \;,\quad \t_n = (-1)^{n-1} \t/2 \;,\quad  \om=-i\b h  \eqn\subs
$$
Notice that the second term in the r.h.s. of \eee\ then vanishes, so that
we have
$$
      {\rm eigenvalue}\{T_N(\t)_h\} = a(\t)^N \; \prod_{r=1}^M
       {{\sinh(\l_j+i\t/2+i\g/2)} \o {\sinh(\l_j+i\t/2-i\g/2)}}
       {{\sinh(\l_j-i\t/2-i\g/2)} \o {\sinh(\l_j-i\t/2+i\g/2)}}     \eqn\eedia
$$
For $0\le\t\le\g$, the largest
eigenvalue $\Lambda_N^{max}(\t)$ of $~T_N(\t)_h~$ is nondegenerate and
correspond to the solution of the BAE with $M=N$ roots which we have called
a.g.s. configuration in sect. 2 . We see therefore that, for $h>0$ ($h<0$)
the roots lay in the lower (upper) complex half--plane and are real for
$h=0$. Moreover, by eq.\symm, $\l_j(h)={\bar\l}_j(-h)$ and
$\l_j=-{\bar\l}_{N-j+1}$. Thanks to the latter symmetry the corresponding
eigenvalue $\Lambda_N^{max}(\t)_h$ remains real, while the former will
ensure that $f(\b,h)=f(\b,-h)$, as required by the spin inversion symmetry
of the XXZ chain. Finally, the nondegeneracy of
$\Lambda_N^{max}(\t)_h$ implies that the two limits in eq.\crossed\
commute [\japs],
so that the limit $L\to\infty$ selects $\Lambda_N^{max}(\g-2\bt/N)_h$
and one finds [\nos]
$$\eqalign{
  f(\b,h) &=-{1\o{2\b}}\lim_{N\to\infty}\log \Lambda_N^{max}(\g-2\bt/N)_h \cr
      &={1\o{2\b}}\lim_{N\to\infty} \left[E_N-
            2N\log{{\sin(2\bt/N)}\o{\sin\g}}\right]   \cr}       \eqn\subtra
$$
where $$\eqalign{
        E_N &=\sum_{j=1}^N \e_0(\l_j)             \cr
        \e_0(\l)&=i\phi(\l_j+i\g/2,\g/2-\bt/N)
        -i\phi(\l_j-i\g/2,\g/2-\bt/N)  \cr}                     \eqn\defi
$$
By the same trick used in sec.2 and 3, we transform the sum over the roots
into integrals and then use eq.\terza\ to eliminate the term linear in
the counting function. We obtain in this way
$$\eqalign{
   E_N = E_c &-i\intf d\l\,  \psi_N(\l+i\eta_+)
        \log\left[1+e^{iZ_N(\l+i\eta_+)} \right]     \cr
         &+i\intf d\l\, \psi_N(\l-i\eta_-)
        \log\left[1+e^{-iZ_N(\l-i\eta_-)} \right]  \cr} \eqn\ener
$$
where $E_c$ is the continuum  approximation
$$
    E_c=N\intf d\l\, \e_0(\l)\s_c(\l)= -2N\intf{{dk}\o k}\,
    {{\sinh(\pi-\g)k\,\sinh(\g-2\bt/N)k}\o{\sinh\pi k\cosh\g k}}   \eqn\econt
$$
and
$$
      \psi_N(\l) = {1\o{2\pi}} [(1-G)\ast \e^\prime_0](\l) = {2\o\g}
         {{\sinh(\pi\l/\g)\,\cos(\pi\bt/\g N)}\o
         {\cosh(2\pi\l/\g)-\cos(2\pi\bt/\g N)}}
$$
Let us recall that the counting function $Z_N(\l)$ fulfills now eq.\terza\
where, in the source term $2\pi Nz_c(\l)$, we must set $\t=\g/2-\bt/N$.
We then see that the root density of the continuum approximation
$$
     \s_c(\l)= {2\o {\g}}{{\cosh(\pi\l/\g)\,\sin(\pi\bt/\g N)}\o
               {\cosh(2\pi\l/\g)-\cos(2\pi\bt/\g N)}}          \eqn\sigl
$$
is strongly peaked at $\l=0$ for large $N$, reflecting
the singularity which develops at the origin in the $N\to\infty$ limit of the
source term in eqs.\linear\ when $\t=\g/2-\bt/N$.
The density picture correctly describes
the BA roots wherever $\s_c(\l)$ is of order 1. From eq.\sigl\ we then
find as validity interval
$$
                       |\l| \leq O(\sqrt{\b/N})                  \eqn\valid
$$
which shrinks to zero when $N\to\infty$. Roots $|\l_j|>O(\sqrt{\b/N})$ have a
spacing of order larger than $O(1/N)$ and cannot be described by densities.
In particular, the roots with largest magnitudes have finite $N\to\infty$
limits spread by $O(1)$ intervals (we checked this fact numerically too).
Therefore, contrary to the usual situation [\dev], we must go beyond the
density description to obtain a bulk quantity like the free energy per site.

We now consider the limit $N\to\infty$; $N$ explicitly
enters eqs.\ener\ and \terza\ only through the functions $\psi_N(\l)$
and $z_c(\l)$. Thus, inserting eqs. \ener\ and \econt\ in eq.\subtra\ yields
$$
      f(\b,h)= E_{XXZ} +\b^{-1}L(\b,h)                       \eqn\separ
$$
where $E_{XXZ}$ is the ground--state energy at zero external field $h$
of \hxxz, namely
$$
     E_{XXZ}=2J\left[\cos\g-\sin\g\, \intf dk\,
       {{\sinh(\pi-\g)k}\o{\sinh\pi k\cosh\g k}}\right]            \eqn\exxz
$$
while (the spin reflection symmetry allows to take the half sum of the two
cases $h\ge0$ and $h\le0$, to ensure explicit reality of the final expression)
$$
        L(\b,h)= \Im \intf d\l \;
        {{\log[1+e^{-\e_+(\l)}][1+e^{-\e_-(\l)}]}
        \o{2 \, \g\, \sinh[\pi(\l+i\eta)/\g]}}          \eqn\correc
$$
The functions $\e_\pm(\l)$ satisfy the limit $N\to\infty$ of eq.\compactpm,
which stays formally unchanged
$$
      \e_\pm = g_\pm - G*\log(1+e^{-\e_\pm}) +
               G_{2\eta}*\log(1+e^{-{\bar\e}_\mp})          \eqn\compactpmi
$$
but where now
$$
        g_\pm(\l) = {{2\pi i\bt}\o{\g\sinh[\pi\l/\g]}}
                        \pm {{\pi\b h}\o{\pi-\g}}               \eqn\newdef
$$
We may also write $\e_\pm(\l)= -iZ(\l+ i\eta;\mp h)$ (compare with
eq.\defeps), with the analytic
function $Z(\l)$ simply related to the original counting function by
$$
  Z(\l)=\lim_{N\to\infty}\left[Z_N(\l)-\pi N{\rm sign\,}(\l)
        \right]     \; \quad (\l\; {\rm real})                    \eqn\relz
$$
The problem of calculating the free energy of the (infinite) XXZ spin chain
has thus been reduced to the problem of solving the two coupled
nonlinear integral equations \compactpmi\ and performing the integral
\correc. Notice that for vanishing as well as for purely imaginary
external field, eq.\defeps\  entails $\e_+=\e_-$, so that eq.\compactpmi\
reduces to a single equation  for $\e(\l) = -iZ(\l+ i\eta)$.

Notice that for $\l \to \pm \infty $, eqs.\count\ and \relz\ yield
$$
Z( \pm \infty ) = - 2 \om
$$
One could take the limit $N\to \infty$ directly in the definition
itself of the counting function, eq.\count, which now reads
$$
   Z_N(\l)= S_N(\l)-\sum_{k=1}^N \phi(\l-\l_k,\g) +2i\b h       \eqn\countI
$$
with the source term
$$
    S_N(\l)=N\left[\phi(\l+i\g/2-i\bt/N,\g/2)+
                   \phi(\l-i\g/2+i\bt/N,\g/2) \right]           \eqn\source
$$
Since the roots $\l_j$ all have finite $N\to\infty$ limits and
$$
    \lim_{N\to\infty} \left[ S_N(\l)- \pi N{\rm sign\,}(\l)
                      -N\phi(\l,\g) \right ] =2\bt q(\l)
\eqn\limsou
$$
with $$
           q(\l)={{\sinh2\l}\o{\cosh2\l-\cos2\g}}- \coth\l        \eqn\defq
$$
we obtain for the limit \relz\ of the counting function
$$
     Z(\l)=2\bt q(\l)-\sum_{j\in\ZZ}\left[\phi(\l-\xi_j,\g)
                              -\phi(\l,\g)\right] +2i\b h    \eqn\discre
$$
where the real numbers $\xi_j$, which satisfy $Z(\xi_j)=(2j-1)\pi$,
$j\in\ZZ$, are the $N\to\infty$ limit of the original BA roots
$\l_1,\l_2,\ldots,\l_N$. Eq.\discre\ shows that $Z(\l)$ has periodicity
$i\pi$ and has, as unique singularity in the strip $|\Im\l|<\g/2$,
a simple pole at the origin with residue $-2\bt$. This fact will play
an important r\^ole in the high temperature expansion of sec. 6.1.

The new BA equations $Z(\xi_j)=(2j-1)\pi$ embody all the information
about the XXZ thermodynamics. They form an infinite set of
algebraic equations equivalent to our non-linear
integral equations \compactpmi . We want to stress that
these roots are {\bf discrete} although we have already set $N =
\infty $. The root spacing is actually of order $T$ and in the $T\to 0$
limit we recover the continuous distribution of roots associated with
the antiferromagnetic ground-state [\dev].
The roots $\xi_j$ have an accumulation point at $\xi = 0$. That is,
$\xi_j = -\bt/[\pi(j-{1/2})]$ for $j\to\infty$. For low temperatures,
the largest root, $ \xi_1 $ , is of order ${(\g/\pi)}\ln\bt$.

We can verify that eq.\discre\ is indeed equivalent to eq.\compactpmi\
by means of the usual contour integrations. Taking into
account that $Z(\l)$ has a simple pole at the origin and using the
identity
$$
 q(\l)= -i\phi^\prime(\l-i\g/2,\g/2)+\frac12 i\phi^\prime(\l,\g)  \eqn\ident
$$
one obtains from eq.\discre\ , for $ \g /2> \Im \l > 0 $ and $h\ge 0$,
$$
   Z(\l) = 2i\bt \; \phi^\prime(\l-i\g/2,\g/2)+ 2i\b h -
            \oint_\Gamma {{d\mu}\o{2\pi i}}\, \phi^\prime(\l-\mu,\g)
            \log[1 + e^{-iZ(\mu)}]                         \eqn\altform
$$
where the contour $\Gamma$ encircles all the roots $\xi_j$ as well as
the origin. By deforming $\Gamma$ into straight lines as in sec. 2,
applying the convolution $(1+K)^{-1}$ and observing that
$$\eqalign{
   \intf d\mu\, (1+K)^{-1}(\l-\mu)\,\phi^\prime(\mu-i\g/2+i0,\g/2) &=
               {{i\pi}\o{\g\sinh[\pi(\l+i0)/\g]}} \cr
  \intf d\mu\, (1+K)^{-1}(\l-\mu) &= {{\pi}\o {2(\pi - \g)}} \cr}
\eqn\identI
$$
one recovers the $N\to\infty$
limit of equation \terza\ (with $\om=-i\b h$ and $\t=\g/2-\bt/N$), from
which eq.\compactpmi\ has been derived.

As we shall see in sec.6.1, the alternative form \altform\ of the integral
equation for $Z(\l)$ is most suitable to study the high temperature
regime, since it involves a complete contour integral. It is
then useful to obtain an analogous formula for the free energy. From
eq.\correc\ we obtain (for $h\ge0$)
$$
        L(\b,h)={1\o2}\intf d\l\, {{Z(\l+i0)}\o{\g\sinh[\pi(\l+i0)/\g]}}
        - {1\o{2i}}\oint_\Gamma d\l\,
        {{\log\left[1+e^{-iZ(\l)}\right]}\o{\g\sinh[\pi\l/\g]}} \eqn\correcI
$$
Substituting eq.\altform\ for the term linear in $Z(\l)$ and using
\identI\ , leads to
$$
        L(\b,h)= \intf d\l\, {{i\bt \phi^\prime(\l-i\g/2+i0,\g/2)+i\b h}
                                \o{\g\sinh[\pi(\l+i0)/\g]}}
        + \oint_\Gamma {{d\l}\o{2\pi}}\, \phi^\prime(\l-i\g,\g/2)
            \log[1 + e^{-iZ(\mu)}]                         \eqn\correcII
$$
By Fourier transformation, the first integral can be rewritten as
$$
        \intf d\l\, {{i\bt \, \phi^\prime(\l-i\g/2+i0,\g/2)+i\b h}
\o  {\g\sinh[\pi(\l+i0)/\g]}}
        =\b \; (2J\cos\g -E_{XXZ} + h ) \eqn\unamas
$$
so that we finally obtain using eq.\separ , \correcII\ and \unamas\
$$
        f(\b,h)=h+2J\cos\g +{{\sin\g}\o\b}\oint_\Gamma {{d\l}\o{2\pi i}}\,
        {{\log\left[ 1 + e^{-iZ(\l)}\right]} \o
        {\sinh\l\,\sinh(\l-i\g)}}                       \eqn\correcIII
$$
It is clear that, to go from eq.\correc\ to eq.\correcII ,
we have just performed in reverse order the same operations that led to
\correc\ from the definitions \subtra\ and \defi. However, in the middle,
we have taken the limit $ N\to\infty $ , which is difficult to take directly
on eqs.\subtra\ and \defi\ . The final result, eq.\correcIII, can
now be transformed, via contour integration, into an infinite sum over
the roots $\xi_j$  with the result

$$
f(\b)=h+2J\cos\g + {1 \o {2 \b}} \log \prod_k (1 + 2 \cos\g)\left[
1 + {{2 \sin \g \sin(s_k - 2 \g)}\o{\cosh(r_k ) - \cos(s_k - 3\g)}}
\right]
$$
where $ r_k = 2 \; \Re(\xi_k) $ and $ s_k = 2 \; \Im(\xi_k) $.

\chapter{Ground state scaling function of the massive Thirring
(sine--Gordon) model}

Let us now reinterpret the diagonal-to-diagonal lattice associated to
the transfer matrix  $ t_{2N}(\l,\t,\om) $ as the unit--time evolution
operator for {\bf Minkowski} spacetime discretized in light-cone
coordinates. That is, the axis correspond to $x\pm t$, ($ x $ and $t$
being the usual space and time variables). For this purpose we choose
$\t=-2i\Th$, with $\Th$ real. Then,  the local $R-$matrices \rma\
are rendered unitary upon multiplication by $a(-2i\Th)^{-1}$,

In this light-cone approach,
we start from the discretized Minkowski 2D space--time
formed by this regular diagonal lattice of right--oriented and
left--oriented straight lines. These represent true world--lines
of  ``bare'' objects (pseudo--particles) which are thus naturally divided in
left-- and right--movers. The right--movers have all the same positive rapidity
$\Th$, while the left--movers have rapidity $-\Th$. One can regard $\Th$ as a
cut--off rapidity, which will be appropriately taken to infinity in the
continuum limit. Furthermore, we shall denote by $V$ the Hilbert space of
states of a pseudo--particle (we restrict here to the case in which $V$ is the
same for both left-- and right--movers and has finite dimension $n =
2$,  although more general situations can be considered [\ddvj,\dev] ).

The dynamics of the model is fixed by the microscopic transition amplitudes
attached to each intersection of a left-- and a right--mover, that is to each
vertex of the lattice. This amplitudes can be collected into linear
operators $R_{ij}$, the local $R-$matrices, acting non--trivially only
on the space $V_i\otimes V_j$ of $i$th and $j$th pseudo--particles.
$R_{ij}$ thus represent the relativistic scatterings of left--movers on
right--movers and depend on the rapidity difference $\Th-(-\Th)=2\Th$,
which is constant throughout the lattice. Moreover,
by space--time translation invariance any other parametric dependence
of $R_{ij}$ must be the
same for all vertices. We see therefore that attached to each vertex
there is a matrix $R(2\Th)^{cd}_{ab}$, where $a,b,c,d$ are labels for
the states of the pseudo--particles on the four links stemming out of
the vertex, and take therefore $n$ distinct values. This is the
general framework of a  vertex model. These local $R-$matrices \rma\
were rendered unitary upon multiplication by $a(-2i\Th)^{-1}$.

The difference with the standard statistical interpretation is
that the Boltzmann \nxl weights are in general complex, since we should
require the unitarity of the matrix $R$. In any case,
the integrability of the model is guaranteed whenever $R(\l)^{cd}_{ab}$
satisfy the Yang--Baxter equations.

For periodic boundary conditions, the one--step
light--cone evolution operators $U_L(\Th)$ and $U_R(\Th)$, which act on the
''bare'' space of states $\H_N=(\otimes V)^{2N}$ , ($N$ is the number of
sites on a row of the lattice, that is the number of diagonal lines),
are built from the local $R-$matrices $R_{ij}$ as [\ddev ] .
$$
         U_R(\Th)= W(\Th)V \;,  \qquad  U_L(\Th) =W(\Th)V^{-1} \; , \quad
         W(\Th)=R_{12}R_{34}\ldots  R_{2N-1\,2N}    \eqn\evol
$$
where $V$ is the one-step space translation to the right.
$U_R$ ( $U_L$ ) evolves states by one step in right (left) light--cone
direction. $U_R$ and $U_L$ commute and their product $U=U_R \, U_L$ is the unit
time evolution operator.

If $\d$ stands for the lattice spacing, the lattice hamiltonian $H$
and  total momentum $P$ are naturally defined through
$$
             U=e^{-i\d H} \;,\qquad  U_R\; U_L^{-1}=e^{i\d P}    \eqn\evolu
$$
These operators directly express in terms of the  alternating transfer
matrix  $ t_{2N}(\l,-i\Th,0) $ as follows
$$
 U_R(\Th)= a(-2i\Th)^{-N}~t_{2N}(-i\Th,-i\Th,0)  ~~,~~
U_L(\Th)= a(2i\Th)^{N}~t_{2N}(+i\Th,-i\Th,0)^{-1}
$$
Therefore, the unit time evolution operator results
$$
        U= e^{-i\d H} = \left[{{a(2i\Th)}\o{a(-2i\Th)}}\right]^N  ~
                t(-i\Th,-i\Th,0)~ t(i\Th,-i\Th,0)^{-1}  \eqn\basicmT
$$
Notice that $H$ and $P$ are (non--local) lattice operators.
For simplicity we set the twist $ \om = 0 $ in this section.
Whenever the $R-$matrix, that is the building block of the
whole construction, satisfies the Yang--Baxter equations, this UV--regularized
QFT is integrable, and will therefore be so also in the continuum limit.
The integrable QFT corresponding to the six--vertex $R-$matrix
given above was identified in
ref. [\ddev] by Jordan--Wigner transforming the local Pauli matrices into
discretized fermionic fields (on the infinite lattice), and then showing that
these fields satisfy a lattice version of the field equation characteristic of
the massive Thirring model. The {\it bare} continuum
fields are recovered when $\d\to 0$ and simultaneously $\Th\to\infty$ with the
bare mass scale $m_0 \sim \d^{-1}\exp(-2\Th)$ held fixed. The
{\it renormalized} ,
physical continuum limit follows instead by keeping fixed the true mass
scale $m \sim \d^{-1}\exp(-\pi\Th/\g)$ ($\Th$ plays essentially the r\^ole of
cutoff in rapidity space). All the on--shell calculations can
be performed exactly within the Algebraic Bethe Ansatz,
leading to the complete spectrum and exact
$S-$matrix of the mT model. Let us also recall that the latter is equivalent
\`a la Coleman--Mandelstam to the sine--Gordon model.
We shall now show what modifications to the general formulae derived in
sect.4 are necessary in order to discuss the thermodynamics of the sG--mT
model.

By the euclidean
symmetry of the corresponding functional integral, we know that the free
energy density of a 2D QFT at temperature $T$ (on an infinite line)
is identical to $T$ times the ground state energy of the same QFT on a
ring of circumference $\b=1/T$. Of course, such a quantity is UV
divergent and must be properly subtracted. Requiring that the free energy
$f(\b)$ vanishes at zero temperature is equivalent to consider only the Casimir
energy, that is the difference between the ground state energy on the
circle and that of the infinite line. In the context of Perturbed Conformal
Field Theory, this Casimir energy is known as ground--state scaling function
$E(\b)$, so that we have $f(\b) = E(\b)$. When $T\to\infty$ we expect that
UV fixed point characterizing the QFT manifests itself in some universal
fashion. In detail, we expect that $E(\b)$ tends to the Casimir
energy of the Conformal Field Theory corresponding to that UV fixed point.
{F}rom this point of view, the situation of a massive QFT such as the
sG--mT model is opposite to that of a gapless spin chain such as the XXZ
model at $0<\g<\pi$. In the former the mass is generated by some IR relevant
perturbation of the CFT and becomes irrelevant in the UV limit $T\to\infty$.
In the latter the lattice cutoff corresponds to the presence of IR irrelevant
perturbations of the CFT those effects become negligible in the $T\to 0$ limit.
Thus the basic properties (central charge and scaling dimensions) of the same
CFT (the free massless boson) will appear at low temperatures in the gapless
XXZ spin chain and at high temperatures in the massive sG--mT model.
This dichotomy will be quite clear in the unified description that we are here
proposing, since, apart from a different source term, the fundamental
integral equation is essentially the same in both cases.

As stated above, the diagonal--to--diagonal transfer matrix $T_N(\t)$,
eq.\tran, once unitarized into $U$ according to eq.\evol, describes the real
time evolution of a discretized mT model. The continuum relativistic QFT, on
the infinite Minkowski plane, is obtained taking first the IR limit
$N\to\infty$ at fixed lattice spacing $\delta$ and then the continuum limit
$\delta\to 0$ near the critical point $\Th=\infty$ (with the physical mass
scale $m \sim \d^{-1}\exp(-\pi\Th/\g)$ held fixed). On the other hand, by
taking
the continuum limit at fixed $N\delta \equiv \b$, we get instead the same QFT
on a ring of length $\b$. In particular, if we consider the ground state on the
lattice, this limit will yield the (UV divergent) bulk ground state energy of
order $\b$ plus the ground state scaling function $E(\b)$.

Using now eq.\enerI\ and \basicmT ,
the lowest value of the energy is then attained in the a.g.s. configuration and
take the form
$$
        E_N =\sum_{j=1}^N \left[ \phi(\l_j-\Th,\g/2)
                -\phi(\l_j+\Th,\g/2) -2\pi\right]               \eqn\defimT
$$
where the BAE roots $\l_j$ are real, since $\om=0$.
In terms of the counting function $Z_N(\l)$, the expression for $E_N$
is identical in structure to eq.\ener, with only $E_c$ and $\psi_N(\l)$
changing to the forms proper of the mT model:
$$\eqalign{
        E_c  &={{N^2}\o\b}\left[-2\pi+\intf d\l\,
           {{\phi(\l+2\Th,\g/2)}\o{\g\cosh\pi\l/\g}}\right]      \cr
  \psi_N(\l) &= {N\o{\g \b}} \left[ {\rm sech\,} \frac\pi\g(\l-\Th)
        -{\rm sech\,}\frac\pi\g(\l+\Th) \right] \cr}            \eqn\mTstuf
$$
In the continuum limit $N\to\infty$ the rapidity cutoff $\Th$ diverges like
$$
        \Th \simeq {\g\o\pi}\log{{4N}\o{m\b}}                   \eqn\runtet
$$
where $m$ is the mass of the mT fermion (or sG soliton). Thus
$$
        E_c = \frac14  m^2\b\,\cot\frac{\pi^2}{2\g}
        + {\rm UV~divergent~terms}                      \eqn\bulkenergy
$$
where the first finite term is the scaling bulk energy (the same result
was obtained in ref.[\dedev] by completely different means). By `UV
divergent terms' we mean terms that scale as the bare mass ($
e^{2\Th}$ ) for large $\Th$. On the other hand
$$\eqalign{
    E(\b) &\equiv \lim_{N\to\infty}\left[ E_N-E_c \right] \cr
          &=-{m\o\g} ~\Im \intf d\l \,
             \sinh\left[\pi(\l+i\eta)/\g \right] \log\left[
            1+e^{iZ(\l+i\eta)}\right]           \cr}            \eqn\gssf
$$
where $Z(\l)$, the $N\to\infty$ limit of the counting function $Z_N(\l)$,
satisfies
$$
    Z(\l)=m\b\,\sinh(\pi\l/\g)+2~\Im \intf d\mu\,G(\l-\mu-i\eta)
        \log\!\left[ 1+e^{iZ(\mu+i\eta)}\right]
\eqn\ntba
$$
Let us recall that this result applies to the complete range $0<\g<\pi$,
provided we take $0<\eta<\frac12 {\rm min}(\g,\pi-\g)$. In the repulsive
regime $\g<\pi/2$ we can cast eqs. \gssf\ and \ntba\ in a very nice form,
reminiscent of the standard TBA. Let us choose $\eta=\g/2$ and set
$$\eqalign{
        & \e_f(\t) = -iZ(\frac\g\pi \t+i\g/2)
                \;,\qquad \e_\fb =\overline{\e_f}       \cr
        & L_f \equiv \log(1+e^{\e_f}) \;,\qquad
                L_\fb \equiv \log(1+e^{\e_\fb}) =  \overline{L_f} \cr
        & G_0(\t)=\frac\g\pi G(\frac\g\pi \t)
                \;,\quad G_1(\t)=G_0(\t+i\pi-i0)   \cr}         \eqn\moredefi
$$
Then the ground state scaling function reads
$$
        E(\b) = -{m\o{2\pi}}\intf d\t\, \cosh \t
                 \left[ L_f(\t)+L_\fb(\t) \right]               \eqn\emT
$$
while the nonlinear integral equation becomes
$$
        \e_f  = g - G_0*L_f + G_1*L_\fb                 \eqn\ours
$$
where now $g(\t)=m\b\cosh\t$ is the source term of the standard TBA, and
is directly related to the level density of free relativistic particles.
Moreover, from eqs. \kernel\ and \moredefi,
$$
        G_0(\t) =\intf{{dk}\o{4\pi}}
        {{\sinh \left( \frac{\pi^2}{2\g}-1 \right)k}\o
        {\sinh \left( \frac{\pi^2}{2\g}-\frac12 \right)k \,
        \cosh  \frac\pi 2 k }} ~e^{ik\t}
$$
By construction, the equation for $\e_\fb$ is just the complex conjugate
of \ours. Let us observe that $G_0(\t)$ is related in the standard way
to the fermion--fermion scattering amplitude $S_{ff}(\t)$
$$
        G_0(\t) = {1\o{2\pi i}} {d\o{d\t}} \log S_{ff}(\t) \eqn\reltosca
$$
while, up to a sign, $G_1(\t)$ coincides with the crossed $G_0$, that is
$G_0(i\pi-\t)$. In other words, we can regard $\e_f$ and $\e_\fb$ as complex
pseudoenergies for the fermions and antifermions, respectively, with
eq.\ours\ as TBA equations involving only physical particles. This situation
should be compared with the standard TBA for the mT (or sG) model [\fozo],
which for nonrational values of $\g/\pi$ involves an infinite number of
nonlinear integral equations for the infinitely many different types of
``magnons" describing the energy degeneracies of the physical fermions and
antifermions. Therefore, {\bf our equation effectively provides
 a resummation of the magnon
degrees of freedom}.

The results \emT\ and \ours\ may be extended to more general situations in
a rather straightforward way. For instance, we could consider a nonzero
twist $\om$ in eqs.\basicmT, or introduce a coupling to the conserved
$U(1)$ charge $Q$ in the real time evolution defined by $U$. Let us consider
first this latter case. The change of the Hamiltonian $H\to H-hQ$, ($h>0$),
has a simple consequence: for $h>m$ the ground state contains a definite
number of fermions (which have charge $Q=1$) with the lowest possible
kinetic energy. On the light--cone lattice this means that the solution of
the BAE must contain holes around $\l=0$, that is roots $\xi_j$ of the
counting function $Z_N(\l)$, $Z_N(\xi_j)= $
which do not appear in the sum over BAE roots in the definition of $Z_N(\l)$.

As in the XXZ chain, we can find for the sG--mT model an infinite set
of algebraic Bethe Ansatz type equations which are equivalent to the
non-linear integral equation \ntba. In order to do that, we define
$$
\nu (\t) \equiv Z(\g\t/\pi) - m\b \sinh(\t)        \eqn\defnu
$$
This function $\nu(\t)$ vanishes at $\t = \pm \infty $. We can then
write,
$$
{\rm Im\,}\log\left[
            1+e^{iZ(\g\t/\pi +i0)}\right]={1\o2}\, \nu(\t) +  \pi D(\t)
                                                        \eqn\regu
$$
where $D(\t)$ is the (subtracted) discontinuity of $\log\left[
            1+e^{iZ(\g\t/\pi +i0)}\right] $ . That is,
$$
D(\t) = {1\o{2\pi i}} \log{{1+e^{iZ(\g\t/\pi +i0)}} \o {1+e^{iZ(\g\t/\pi-i0)}}}
 + { m\b \o {2\pi}} \sinh (\t)
                                                        \eqn\ddis
$$
Then,
$$
D'(\t) = {m\b \o {2\pi}} \cosh(\t) -\sum_{k}\delta (\t - \t_k)    \eqn\dpz
$$
where the $\t_k$ fulfil the equations:
$$
Z(\g\t_k/\pi) = (2k+1)\pi \, , \qquad k \,\e \, Z
                                                \eqn\bat
$$
Since $Z(-\t)= - Z(\t)$,  the $\t_k$ are symmetrically
distributed with respect to the origin:
$$
\t_k = - \t_{-k-1}
                                                \eqn\dila
$$
Using the asymptotic behaviour $ Z(\g\t/\pi)\buildrel{\t \to\infty}\over =
m\b \, \sinh(\t)$ , we find for large roots $\t_k$ ,
$$
\t_k \buildrel{k\to\pm\infty}\over\simeq \pm\log \left[{{2\pi}\o{m\b}}|2k+1|
\right]
                                                \eqn\asil
$$
We see from eq.\asil\ that the term in $\cosh\t$ in eq.\dpz\ simply
ensures the finiteness of $D'(\t)$. As eq.\asil\ shows, the roots
accumulate at infinity in the mT-sG model, whereas they accumulate at
the origin for the XXZ chain at temperature $T$ (sec.4).

Eq.\ntba\ for $ \eta \to 0^+ $
can be now written as
$$
\nu(\t) = \intf d\mu ~G_0(\t - \mu) \left[ \nu(\mu) + 2\pi D(\mu) \right]
                                                \eqn\nunu
$$
Upon Fourier transform and pulling $\nu(\t)$ to the l.h.s., this yields
$$
\nu(\t) = ({\g \o \pi} )^2\;  \intf d\mu ~\phip(({\g \o
\pi})^2\left[\t-\mu\right],\g)\; D(\mu)
\eqn\inu
$$
or
$$
Z(\g\t/\pi) = m\b \sinh \t + \intf d\mu ~\phi(({\g \o
\pi})^2\left[\t-\mu\right] ,\g)\; D^{\prime}(\mu)
                                        \eqn\eqe
$$
This is the analog of eq.\altform\ . Inserting now eq.\dpz\ into
eq.\eqe\ we find for $ \g > \pi /2$
$$
Z(\g\t/\pi) = - \sum_{k}\phi(({\g \o \pi})^2\left[\t-\t_k \right],\g)
                                        \eqn\epsl
$$
where the sum is to be understood in principal part.
This shows explicitly that $Z(\l)$ has ${i\pi^2 \o \g}$ as
period. Moreover, setting $ \t = \t_l$ in eq.\epsl\ and using eq.\bat\ ,
we find
$$
(2l+1)\pi = - \sum_{k}
\phi(({\g \o  \pi})^2\left[\t_l-\t_k \right],\g)
                                        \eqn\baeft
$$
always for $ \g > \pi/2$.

\chapter{The Recursive Regime}

Depending on the temperature regime chosen, the integral equations
\altform\ and \ours\ can be simply solved by iterating the
inhomogeneity.
This happens for small $\b$ in the XXZ chain [eq.\altform ] and for
large   $\b$ in field theory [eq.\ours\ for the mTm ]. We call
`recursive regimes' those where such expansions apply.

\section{High Temperatures in the XXZ chain}

We study here the free energy $f(\b)$ of the XXZ chain
for high temperatures. When $\b$ is small
it is convenient to use the form   \altform\ plus the uniform
expansion$$
            Z(\l)=\sum_{k=1}^\infty \bt^k\,b_k(\l)              \eqn\unif
$$
Since the residue of $Z(\l)$ is linear in $\bt$ only , $ b_1(\l)$ has
a pole at the origin, we have
$$
          b_1(\l)\buildrel{\l\to0}\over\simeq -{2\o{\l}}+O(\l)   \eqn\bprop
$$
while all the $b_k(\l)$ for $k \ge 2$ are analytic there.

{F}rom eq.\unif\ we find up to  fourth order in $\bt$
$$\eqalign{
\log\left[1 + e^{-i Z(\l)} \right] = & \ln 2 -
{{i \bt} \o 2} \; b_1(\l) \cr
& -{{\bt^2} \o 2} \left[ i b_2(\l) + {1 \o 4} b_1(\l)^2 \right] -
{{\bt^3} \o 2 } \left[ i b_3(\l) + {1 \o 2} \; b_1(\l)\;
b_2(\l)\right] \cr
 & - {{\bt^4} \o 2 } \left[  i b_4(\l) + {1 \o 2} \; b_1(\l)\;
b_3(\l) + {1 \o 4} b_2(\l)^2  + {1 \o 96} b_1(\l)^4 \right] + O( \bt^5) \cr}
\eqn\explo
$$
We insert this expansion into eq.\altform\ and notice that only the
poles at $\l = 0$ in eq.\explo\ contribute to the contour integral
over $\Gamma$ . Since only $b_1(\l)$ is singular, the procedure is
{\bf perfectly recursive} and we find
$$
\eqalign{
  b_1(\l) &= 2 q(\l)+ 2i\ht \cr b_2(\l) &=
- {1\o{2}}\phi^{\prime\prime}(\l,\g) -  i\ht\phi^{\prime}(\l,\g)\cr
      b_3(\l) &= i\ht \phi^{\prime}(\l,\g) \cot\g  \cr
      b_4(\l) &= -i\ht\left({1\o6}+{1\o{2{\rm sin}^2\,\g}}-{{\ht^2}\o3}\right)
                \phi^{\prime}(\l,\g) +
 {1\o 3}\left(-{1\o 3}+{1\o{{\rm sin}^2\,\g}} + {3 \o 2} \ht^2\right)
                \phi^{\prime\prime}(\l,\g)   \cr
    &-{{i\ht}\o 6}\phi^{\prime\prime\prime}(\l,\g)
      -{1 \o {72}} \phi^{\prime\prime\prime\prime}(\l,\g) \cr } \eqn\fourth
$$
where $\ht = h/(J\sin\g)$. Next we insert the expansion \explo,
with the explicit results listed in \fourth, into the formula \correcIII\
of the free energy. Again all we need to do is to repeatedly apply
the residue theorem, with the final result
$$\eqalign{
       f(\b)= &-\b^{-1}\ln2 +J\cos\g-\b \left[
         J^2(1+\frac12{\rm cos}^2\,\g)+\frac12 h^2 \right]
                +\b^2\cos\g (J^3+h^2) \cr
        &- \b^3 [(\frac14 +\frac16 \cos^2\g -\frac1{12}\cos^4\g)J^4 -
                J^2 h^2 -\frac1{12} h^4 ] +O(\b^4)  \cr} \eqn\third
$$
which indeed agrees with the high $T$ expansion, as can be derived direclty
from the definitions \free\ and \hxxz.

\section{Low Temperatures in field theory}

For low temperatures (large $m\b$) the non-linear integral equation
\ours\
valid for $\g < \pi /2$,
$$\eqalign{
\e_f(\l)=&~ m \b\cosh \l~ + \cr
& \intf d\mu\left\{ G_1(\l-\mu ) \log\!\left[
            1+e^{-{\bar\e}_f(\mu)}\right] - G_0(\l-\mu)\log\!\left[
            1+e^{-\e_f(\mu)}\right] \right\} \cr}
\eqn\nova
$$
can be easily iterated using the inhomogeneity
$m \b\cosh \l$ as zeroth order approximation. That is,
$$\eqalign{
\e_f(\l) =& ~f_0(\l) + f_1(\l) + f_2(\l) + \ldots \cr
f_0(\l) =& ~ m \b \cosh \l \cr
 f_1(\l) =& \intf d\mu\left[\; G_1(\l-\mu )\,
- G_0(\l-\mu)\, \right]\; \log\! \left[ 1 + e^{-m \b\cosh \mu} \right] \cr
 f_2(\l) =&  \int_{-\infty}^{+\infty}{{d\mu} \o {e^{f_0(\mu)} + 1}}
\left[ G_0(\l-\mu) \; f_1(\mu)\,
- G_1(\l-\mu)\; \overline{f_1(\mu)}\, \right] \; \cr}
\eqn\fs
$$
The energy $E(\b)$ is expanded in analogous way as a sum:
$$
E(\b) = E_0(\b) + E_1(\b) + E_2 (\b)+ \ldots
\eqn\es
$$
where
$$\eqalign{
E_0(\b) =& - { m \o {\pi}} \int_{-\infty}^{+\infty}d\mu\; \cosh \mu
\; \log\! \left[ 1 + e^{-m\b \cosh \mu} \right] \cr
E_1(\b) =& ~ { m \o {2 \pi}} \int_{-\infty}^{+\infty}d\mu\; \cosh \mu \;
 e^{-f_0(\mu)} \left[ f_1(\mu) + \overline{f_1(\mu)} \right] \cr}
\eqn\eu
$$
Let us now analyze more explicitly $E_0(\b)$ and $E_1(\b)$. From eq.\eu\ we
find
$$
 E_0 (\b)= -{{2m}\o\pi} \sum_{n=1}^{\infty} {{(-1)^{n-1}} \o n} K_1(nm\b)
\eqn\ece
$$
This is the typical behaviour of a massive QFT.
Notice that eq.\ece\ for $\g = \pi/2$ gives the {\bf exact} energy $E(\b)$.
(In that free case $G_0(\l) = 0$ ).

For large $m\b$, eq.\ece\ yields
$$
 E_0(\b) \buildrel{\b\to\infty}\over=  -{{2m}\o\pi}K_1(m\b)+ O(e^{-2m\b})
\buildrel{\b\to\infty}\over=
-\sqrt{{{2m}\o{\pi \b}}}~e^{-m\b}~\left[ 1 + O({1\o \b}) \right] +
O(e^{-2m\b})
\eqn\eoas
$$
It is easy to see that $E_n(\b)$ is of order $e^{-(n+1)m\b}$ for large
$m\b$.

We get from eqs.\fs\ and \eu\
$$\eqalign{
E_1(\b) =&  -{ m \o {2 \pi}} \int_{-\infty}^{+\infty}d\l ~ \cosh \l ~
 e^{-m\b \cosh \l} \int_{-\infty}^{+\infty}d\mu ~
\log \left[ 1 + e^{-m\b \cosh \mu} \right]  \cr
 & \left[\; 2\, G_0(\l - \mu) -  G_1(\l-\mu )- {\bar G_1(\l-\mu ) }
\; \right]
\cr}
\eqn\eui
$$
The terms inside braces $[\ldots]$ in eq.\eui\ admit for $\g < \pi/2$
an integral representation that follows from eq.\kernel\ and eq.\moredefi\
$$
2\, G_0(\l ) -  G_1(\l ) - {\bar G_1(\l) }=-\int_{-\infty}^{+\infty}
{{dx} \o {2\pi}} ~e^{ix\l}~[ 1 - g({{\pi x} \o 2}) ]
\eqn\gr
$$
where
$$
g(y) = { { \sinh(4 - {{\pi}\o {\g}})y + 3 \sinh({{\pi}\o {\g}}-2 )y }
\o { 2 \; \cosh y \; \sinh({{\pi}\o {\g}}-1 )y }}
\eqn\gi
$$
For large $m\b$ we can approximate the $\log$ in eq.\eui\ by the first
term in its power  expansion. This yields,
$$\eqalign{
E_1(\b) = & -{ m \o {2 \pi}} \int_{-\infty}^{+\infty}d\l \, \cosh \l \;
 e^{-m\b \cosh \l} \int_{-\infty}^{+\infty}d\mu \;  e^{-m\b \cosh \mu}
  \int_{-\infty}^{+\infty}
{{dx} \o {2\pi}} ~e^{ix(\l-\mu)}~[ 1 - g({{\pi x} \o 2}) ] \;
 \;  \cr
     & + O(e^{-3m\b}) \cr}
\eqn\eua
$$
This integral can be approximated by the saddle point method for large $m\b$
with the following result:
$$
E_1(\b)\buildrel{\b\to\infty}\over= e^{-2m\b}\left[ \sqrt{{m\o{2\pi \b}}}
+{{K_{\g}}\o \b} + O({1\o {\b^{3/2}}}) \right]  + O(e^{-3m\b})
\eqn\euas
$$
where
$$
K_{\g} = - \int_{-\infty}^{+\infty} {{dy} \o {\pi^2}}~g(y) =
-\int_{-\infty}^{+\infty} {{dy} \o {\pi^2}}~
 { { \sinh(4 - {{\pi}\o {\g}})y + 3 \sinh({{\pi}\o {\g}}-2 )y }
\o { 2 \; \cosh y \; \sinh({{\pi}\o {\g}}-1 )y }}
\eqn\kag
$$
This function of $\g$ can be expressed as an infinite sum of $\psi$
functions. For rational values of $\g/\pi$ it admits simpler
expressions. For example,
$$
K_0 = - {1 \o {\pi^2}}(4 \ln 2 - 1 ) \quad , \quad K_{\pi/3} = -{2\o{\pi^2}}
$$
In addition, we see from eq.\kag\ that $ K_{\g} < 0 $ for $\g < \pi/2
$ and that  $\lim_{\g \to \pi/2^-} K_{\g} = - \infty $. This
divergence indicates a change of regime since $E_1(\b)$ is in fact
identically zero at $\g = \pi/2$.

In summary,  $E_n(\b)$  for large
$m\b$ is at leading  order $O(e^{-(n+1)m\b})$
 and  contains contributions with arbitrary higher  powers of $e^{-m\b}$.

{F}rom $E_0(\b) + E_1(\b)$ we can write $E(\b)$ up to contributions $
O(e^{-3m\b})$. We find from eqs. \euas\ and \eoas\ ,
$$
E(\b) \buildrel{\b\to\infty}\over= -{{2m}\o\pi}K_1(m\b)+
 e^{-2m\b}       \sqrt{{m\o{\pi \b}}} \left[ {{1+\sqrt 2}\o 2}
+\sqrt{{\pi}\o{m\b}}~ K_{\g} + O({1\o \b}) \right]  + O(e^{-3m\b})
$$
The full term of order $e^{-2m\b}$ in $E_1(\b)$ possess the following
integral representation:
$$\eqalign{
E_1(\b) \buildrel{\b\to\infty}\over=&  {m \o
{\pi}}\int_{-\infty}^{+\infty}dx ~ K_{ix}(m\b) \left[ K_{1+ix}(m\b) +
K_{1-ix}(m\b) \right] \cr
 & {{\sinh^2(\pi x/2) \sinh\left(\pi x [{{\pi}\o {2\g}} -
1 ]\right) } \o{\cosh(\pi x/2) \sinh\left(\pi x [{{\pi}\o {\g}} -
1 ]\right) }} +   O(e^{-3m\b}) \cr}
$$
where we used eqs.\eui , \gr\ and \gi\ .

\chapter{The conformal regime}

The conformal regime is the opposite to the recursive regime. That is,
large $\b$ for magnetic chains and small  $\b$ in field theory.

In the conformal regime the calculations are more subtle but still
straightforward as shown below.

The starting point is again the integral equation \terza . For
simplicity  we shall consider here only real
twists $\om$, that is purely imaginary external fields $h$ for the XXZ chain.
Then, $Z(\l) = {\bar Z({\bar \l}) }$ and we can write for all cases
$$
-i \log F(\mu) = \varphi (\mu) + 2\; \Im \intf d\,\mu' \;G(\mu - \mu' -
i\eta) \log\left[1 + F(\mu' +i\eta) \right]
\eqn\univ
$$
where $ F(\mu) = e^{iZ(\mu)} $ and
$$
         \varphi (\mu) = \cases{
        2N\arctan \left[{{\sinh(\pi\mu/\g)}\o{\cos(\pi\t/[2\g])}}\right]
        -{{\pi\om}\o{\pi-\g}}   & {\rm finite size vertex model} \cr
        -{{2\pi \bt}\o{\g\sinh[\pi\mu/\g]}} +  {{i \pi\b h}\o{\pi-\g}}
                                & {\rm XXZ thermodynamics}      \cr
         m\b\, \sinh(\pi\mu/\g) &  {\rm sine-Gordon field theory} \cr}
                                                          \eqn\funiv
$$
The solution  $Z(\l)$ must be then inserted in eq.\fscorr .
[ Recall that $\e(\mu) = -iZ(\mu)$] .

\section{Leading finite size corrections for the six-vertex model}

Let us derive here the dominant finite size corrections for the
six-vertex model free energy [see sec. 3].

We see from eq.\funiv\ that $\varphi (\mu) \sim N $ for large $N$
 as long as $ \mu \leq (\g / \pi ) \log[ N \cos({{\pi \t}\o{\g}})] $.
Hence,   $ Z(\mu)\sim N $ and we find that this bulk contribution to $
L_N $ vanishes to the order $N^0$ and $N^{-1}$.

The contributions  to $L_N $ of order $N^{-1}$ (and smaller) come from
values of $|\mu|$ larger than $(\g/\pi)\log[ N \cos({{\pi \t}\o{\g}})]$,
where the previous estimate  $ Z(\mu)\sim N $
does not hold anymore.
In order to compute the contributions from large positive values of $\mu$
one introduces the new function
$$
         F(x)=e^{-\e(\mu)}\;,\quad x=\mu-{\g\o\pi}
\log[4 N \cos({{\pi \t}\o{\g}})] \eqn\nvax
$$
Then eqs. \univ\ and \fscorr\ reduce, in the $N\to\infty$ limit, to
$$
-i \log F(x) = - e^{-{{\pi x}\o {\g}}} -{{2\pi\om}\o{\pi-\g}}
\; \Im \intf d\,y \;G(x - y -
i\eta) \log\left[1 + F(y +i\eta) \right] \eqn\unias
$$
$$
L_{2N} = i \om - {{i\;e^{-{{i\pi\l}\o{\g}}}}\o
{N \; \g \; \cos({{i\pi\l}\o{\g}})}}
\intf dx ~e^{-{{\pi x}\o{\g}}} ~ \Im\left\{ e^{-{{i\pi\eta}\o{\g}}}~\log
[1 + F(x + i \eta) ] \right\}
\eqn\cocon
$$
For simplicity, we shall choose $ \eta = 0^+ $ and use that
$G(x+i0) = G(x)$ and $G(\pm \infty) = 0$ for real $x$.
We shall analogously compute below   the contributions from large
negatives  values of $\mu$.

Fortunately, it is not necessary to solve the integral equation \unias\
in order to calculate
the integral \cocon . In fact, we can use the following lemma

Lemma. Assume that $F(x)$ satisfies the nonlinear integral equation
$$
   -i\log F(x)=\varphi(x)+ 2  \int_{x_1}^{x_2}
                  dy\,G(x-y)\, \Im \log[1+F(y+i0)]          \eqn\nlnr
$$
where $\varphi(x)$ is real for real $x$ and $x_1,\,x_2$ are real numbers.
Then the following relation holds
$$\eqalign{
    \Im &\int_{x_1}^{x_2}dx\,\varphi^\prime(x)\log[1+F(x+i0)]= \cr
   \frac12 &\;\Im
\left[\varphi(x_2)\log(1+F_2)-\varphi(x_1)\log(1+F_1)\right]
             +\frac12 \; {\rm Re\,}\left[\ell(F_1)-\ell(F_2)\right] \cr
     +&  \int_{x_1}^{x_2}dy
\left[ \, G(x_2 -y) \log(1+F_2) -  G(x_1 -y) \log(1+F_1) \, \right]\;
\Im \log[1+F(y+i0)]
\cr} \eqn\lemm
$$
where $F_{1,2}=F(x_{1,2})$ and $\ell(t)$ is a dilogarithm function
$$
      \ell(t)\equiv\int_0^t du\,
\left[{{\log(1+u)}\o u}-{{\log u}\o{1+u}}\right]    \eqn\dilog
$$
To prove the lemma we consider the relation
$$     \ell(F_2)-\ell(F_1)=
       \int_{x_1}^{x_2}dx\left\{\log[1+F(x+i0)]{d\o{dx}}\log F(x) -
                  \log F(x){d\o{dx}}\log[1+F(x+i0)]\right\}
$$
and then use eq.\nlnr\ and its derivative to substitute $\log F(x)$
and $d\log F(x)/dx$. This yields
$$\eqalign{
\ell(F_2)-\ell(F_1)=&\; i \int_{x_1}^{x_2}dx\left\{ {{d\varphi(x)}\o {dx}}
\log[1+F(x+i0)] - \varphi(x){d\o{dx}}\log [1+F(x+i0)] \right. \cr
&  + \log [1+F(x+i0)]\int_{x_1}^{x_2}dy\; G'(x-y)\;\Im
\log[1+F(y+i0)] \cr
-& \left. {d\o{dx}}\log [1+F(x+i0)]
\int_{x_1}^{x_2}dy\; G(x-y)\;\Im \log[1+F(y+i0)]
\right\} \cr}
$$
The double integral here cancels upon partial integration and using
that $ G'(x-y) = - G'(y-x) $. Finally, taking real part  yields
eq.\lemm\ (related identities were used in ref. [\aus]).

We can now use the lemma to compute the integral \cocon\
setting
$$
\varphi(x) =  - e^{-{{\pi x}\o {\g}}} -{{\pi\om}\o{\pi-\g}}~~
{\rm and} ~~ x_1=-\infty~,~
x_2=+\infty.
$$
 We have $ F(x_1) = 0 $, $ F(x_2) = e^{-2i\om} $ and
$$
{{\pi}\o{\g}}\intf dx ~ e^{-{{\pi x}\o {\g}}}\log[1 + F(x+i0)] =
-{1\o 4}\left[ \ell( e^{-2i\om} ) + \ell( e^{+2i\om} )\right] +
{{\om^2} \o {2(1 - \g/\pi)}} \eqn\aple
$$
In this expression we can apply the formula
$$
\ell(z) + \ell(1/z) = {{\pi^2}\o 3} \eqn\iddi
$$
The contribution  from large negatives  values of $\mu$ is
computed analogously introducing the variable
$$
 x' =\mu+{\g\o\pi} \log\left[4 N \cos(\frac{\pi\t}{\g})\right]
$$
We find from eqs. \cocon\ ,\aple\ and \iddi\ collecting both contributions :
$$
L_{2N} = {{\pi}\o{6N}} ~ \tan\left({{\pi \l}\o {\g}}\right)~
\left[1 - {{6\om^2}\o {\pi^2 (1 - \g/\pi)}} \right] \eqn\carc
$$
Hence the central charge turns to be unit.
The factor $\tan\left({{\pi \l}\o {\g}}\right)$ is a geometric effect
due to the kind of diagonal-to-diagonal lattice we are using.
(In the row-to-row framework a factor $\sin\left({{\pi \l}\o {\g}}\right)$
arises [\dev]).

Let us prove that this $\tan$ factor is just the speed
of sound for low-lying excitations. The eigenvalue of $\log{t_{2N}(\l,\t,\om)}$
for a hole excitation normalized to the a.g.s. writes [\dev]
$$
-2i\arctan\left[e^{{{\pi}\o{\g}}(\phi+i\l)}\right] \eqn\buco
$$
where $\phi$ stands for the position of the hole.
Since we are here on an euclidean lattice, we can write $
\log{t_{2N}(\l,\t,\om)} = h + ip $ to define the time and space
generators $h$ and $p$, respectively.
For large positive or negative  $\phi$, the eigenvalues of  $h$ and
$p$ result from eq.\buco\
$$
\e = \pm 2 \sin\left({{\pi\l}\o{\g}}\right)~e^{-{{\pi|\phi|}\o{\g}}} ~~,~~
p =  \pm 2 \cos\left({{\pi\l}\o{\g}}\right)~e^{-{{\pi|\phi|}\o{\g}}}
$$
That is, the speed of sound turns out to be precisely $\tan\left({{\pi
\l}\o {\g}}\right)$.

This result can be interpreted as a proof of the presence of conformal
invariance.

\section{Low temperatures in the XXZ chain}

We shall consider now the low temperature regime. When $\b\gg1$ eq.
\compactpmi\
indicates that $Z(\l)\sim\b$, so that $\log[1+e^{iZ(\l)}]\simeq  iZ(\l)$,
at least as long as $|\l|\leq (\g/\pi)\log\b$. At dominant $\b\gg1$ order
eq.\compactpmi\ linearizes in the same way as it does for small $\b$.
Inserting then $Z(\l)\simeq 2 \bt q(\l)$ in eq.\correc, yields
$\lim_{\b\to\infty}L(\b)=0$. Therefore eq.\separ\ tells us that
$$
     f(\b)\buildrel {\b\to\infty} \over = E_{XXZ}(\g)+O(\b^{-2})\eqn\smallb
$$
The contributions of order $\b^{-2}$ (and smaller) come from values of $\l$
larger than $(\g/\pi)\log\b$, where the previous assumption $\log a\sim O(\b)$
does not hold anymore. It is then convenient to introduce the new function
$$
         F(x)=e^{iZ(\l)} \;,\quad x=\l-{\g\o\pi}\log{{4\pi\bt}\o\g}
\eqn\resc
$$
Then eqs. \correc\ and \compactpmi\ reduce, in the $\b\to\infty$ limit, to
$$
    L(\b)={1\o{\pi\bt}}\intf dx\,e^{-\pi x/\g}~\Im
                                             \log[1+F(x+i0)]  \eqn\cortwo
$$
and$$
        -i\log F(x)= -e^{-\pi x/\g}
        +2\intf dy\,G(x-y)\,\Im \log[1+F(y+i0)]          \eqn\nontwo
$$

The integral in eq.\cortwo\ may now be exactly calculated by using  the
lemma from section 7.1 with  $\varphi=-\exp(-\pi x/\g)$ and $x_1=-\infty$,
$x_2=+\infty$. We have $F(x_1)=0$, $F(x_2)=1$ and
$$
     2\,\Im \intf dx\,e^{-\pi x/\g}~\log[1+F(x+i0)]=
                               -{\g\o\pi}\ell(1) =-{{\pi\g}\o 6}
$$
Then the free energy for low temperature reads
$$
        f(\b)=E_{XXZ}(\g)-{\g\o{12 J\sin\g}}\b^{-2} +o(\b^{-2})   \eqn\lowt
$$
in perfect agreement with refs. [\tak,\devw,\devk].

As we have just shown, both the
high and the low temperature leading behaviors of the free energy can be
derived without much effort from our non--linear integral equation \compact.
The higher order corrections for high $T$ can be obtained in a very systematic
way. The situation for the $o(\b^{-2})$ terms in the low $T$
expansion \lowt\ is more involved.

However, the NLIE \compactpmi\ admits a Riemann--Hilbert formulation
analogous to the one presented in sec. 7.4 for the sG-mT model. One
can establish from such Riemann--Hilbert formulation the analytic
structure of the low $T$ expansion.

\section{High temperatures in the mT-sG field theory}

Let us recall the integral equation \ours\
$$
        \e = r\,ch  -G_0*L + G_1*{\bar L}               \eqn\oursI
$$
where $\e\equiv\e_f$, $r \equiv m \b $ and $ch$ stands for the hyperbolic
cosine, $ch(\t)=\cosh\t$.

For small $r$ (high temperatures), the regions with positive and
negative $\t$ where $ r \cosh \t \sim 1 $ are very far apart. It is
then useful to separately study eq.\ours\ with inhomogeneities
$$
{r \o 2 } e^{\pm\t} = e^{\pm\t -\log 2/r}~.
$$
That is, we define the kink function $\e_k$
as the solution of the same integral equation with $r\,ch$ replaced
by an exponential
$$
        \e_k =  e^\t -G_0*L_k + G_1*{\bar L}_k  \eqn\kink
$$
where $L_k=\log(1+e^{-\e_k}) $. (Analogous kink functions were first
considered in [\albz] and have now become standard in TBA calculations).

Then, we set
$$\eqalign{
        \eta(\t) &=\e(\t)-\e_k(\t-\log2/r)-\e_k(-\t-\log2/r)  \cr
        \ell(\t) &=L(\t)-L_k(\t-\log2/r)-L_k(-\t-\log2/r)+\log2 \cr} \eqn\kkk
$$
{F}rom eqs. \oursI\ and \kink, using the translational invariance of
eq.\kink\ $\eta$ and $\ell$ turn to be  related by an
homogeneous equation
$$
        \eta = -G_0 * \ell +G_1 *\ell                   \eqn\etaell
$$
We then find for the energy
$$\eqalign{
        E(\b) &=-{m\o{2\pi}}\intf d\t\,\cosh\t
                \left[L(\t)+ {\bar L}(\t)\right] \cr
          &=-{m\o{2\pi}}\left (f_k+{\bar f}_k \right)
            -{m\o{2\pi}}\intf d\t\,\cosh\t
        \left[\ell(\t)+{\bar\ell}(\t)\right] \cr } \eqn\enerV
$$
where the kink contribution reads
$$\eqalign{
        f_k &= \intf d\t\, e^\t \left[
                L_k(\t-\log2/r)+L_k(-\t-\log2/r)-\log2 \right] \cr
            &= {2\o r}\intf  d\t\, e^\t  L_k(\t) +
        \intf  d\t\ e^{-\t} \partder~\t \left[L_k(\t-\log2/r)-\log2\right] \cr
            &= {2\o r}\intf d\t\, e^\t  L_k(\t) +
        {r\o2}\intf d\t\, e^{-\t} ~\partder{L_k}\t \cr} \eqn\morekink
$$
The real parts of these two integrals can be calculated without
explicit knowledge of the kink function $\e(\t)$.
To the term proportional to $r^{-1}$ we can apply the Lemma with the
substitutions $\varphi(\t)=e^\t$ and $x_1=-\infty$, $x_2=+\infty$.
This gives
$$
        \intf d\t\, e^{\t}\, \left[ L_k(\t)+ {\bar L}_k(\t) \right]
        = \ell(1) = {{\pi^2}\o6}                        \eqn\cpart
$$
To compute  the term linear in $r$ in eq.\morekink\ we can adapt to our
situation an argument in ref.[\albz ] : to converge for
$\t\to -\infty$, the second integral
requires that $\partial{L_k}/\partial\t$ decays faster than $e^\t$ , hence
also $\e_k(\t)$ must tend to $0$ faster than $e^\t$ as $\t\to -\infty$.
We then differentiate eq.\kink\  with respect to $\t$, let
$\t\to -\infty$ and use the asymptotic expansion for large $|\l|$
$$
        G_0(\l) \simeq a_1 \, e^{-|\l|} \left[1+O(e^{-2|\l|})\right] +
        b_1 \, e^{-2\gt|\l|/\pi} \left[1+O(e^{-2\gt|\l|/\pi})\right]
$$
where $\gt=\g\;(1-\g/\pi)^{-1}$ and
$$
        a_1={1\o\pi}\tan \left({{\pi^2}\o{2\g}}\right) \;,\quad
        b_1={\gt\o{\pi^2}}\tan \left({{\pi^2}\o{\pi-\g}}\right)
$$
Thus we formally obtain for $\t\to -\infty$
$$
        {{d\e_k}\o{d\t}} \simeq \left[1 - a_1\intf d\l\, e^{-\l}
        \left( \partder{L_k}\l + \partder{{\bar L}_k}\l \right) \right]e^\t
        + O(e^{2\t}) + O(e^{2\gt\t/\pi})                \eqn\bulkpart
$$
Whether $e^{2\t}$ or $e^{2\gt\t/\pi}$ is the dominant order depends on
who is larger between $\g$ and $\pi/3$, but it does really affect our
argument. In any case self--consistency requires that the coefficient
of $(e^\t)$ vanishes, yielding the value of the real part of the
second integral in eq.\morekink.

Collecting the results of eqs.\cpart\ and \bulkpart\ into eq.\enerV\
one obtains in explicit form the kink contribution to the energy
$$
        E\buildrel {\b\to0} \over = -{{\pi}\o{6\b}}
        - \frac14 m^2\b\,\cot\frac{\pi^2}{2\g}
        -{m\o{2\pi}}\intf d\t\,\cosh\t
        \left[\ell(\t)+\bar{\ell(\t)}\right]            \eqn\enerVI
$$
The first term is the universal conformal Casimir
energy, from which we read the correct central charge $c=1$ of the mT-sG
model.  The second term, linear in $\b$, exactly coincides with
minus the scaling bulk free energy (see eq.\bulkenergy). As is well
known, this is not an accident, since the UV--independent part of the
complete ground state energy $E_c+E \simeq \lim_{N\to\infty} E_N$ for
small $r$ can be calculated through Conformal Perturbation Theory
(PCT). For the mT model the unperturbed conformal theory is the
massless Thirring model, while the perturbation is the mass term. This
has noninteger dimension for $\g\ne\pi/2$, yielding only noninteger
powers of $r$ in the perturbation series for the free energy, apart
from the conformal $\b^{-1}$ term. Hence no term linear in $r$ may
appear. This argument has been often used to compute the scaling bulk
energy of Perturbed Conformal Models from the TBA derived from the
corresponding scattering theory. In our case we verify the validity of
the argument since the scaling bulk energy can be directly calculated
from the microscopic BA solution.

The last integral in eq.\enerVI\  represents the resummation of the PCT.
The fact that it must contain non-integral
powers of $r$, causing non-analyticity at $r=0$,
can be established directly from the original equation
\oursI, since this cannot be expanded in a Taylor series of $r$.
To establish the absence  of any integral power of $r$, and
especially of the first power, without referring to PCT, a more
elaborate treatment of eq.\oursI\  is necessary. This shall be the
subject of next section.

\section{ The NLIE as a Riemann--Hilbert (or Wiener--Hopf) problem.}

In this section we recast the  non-linear integral equations \ntba\
for the mT-sG model as a Riemann-Hilbert (or Wiener-Hopf) problem.
The XXZ thermodynamics can be treated analogously. The techniques
involved are rather standard, having been used very often to study
Bethe Ansatz systems coupled to external fields at zero temperature
(see for instance [\pwr ]). Here we adapt them
to the regime $T>0$ at zero external field.

Let us define,
$$
\rho(\t) \equiv {1 \o {2\pi}} {d\o{d\t}} Z(\g\t/\pi)
\eqn\densi
$$
For large $\b$, the roots $\t_k$ defined by eq.\bat\ become a continuous
distribution with a density precisely given by $\rho(\t)$. For
arbitrary real values of $\b$,  $\rho(\t)$ is just a continuous
function that we are interested to find.
 From eqs.\defnu\ and \densi\ we find
$$
{{d\nu}\o{d\t}}= 2\pi \rho(\t) - m\b \cosh\t
\eqn\denu
$$
This equation combined with the $\t$ derivative of eq.\nunu\ and with eq.\dpz\
yields upon partial integration
$$
 \rho(\t) - {{m\b}\o{2\pi}} \cosh\t = \intf d\mu \;
G_0(\t-\mu)\;\left[\rho(\mu) - \sum_{k\in\ZZ}\delta(\mu - \t_k)\right]
\eqn\ecro
$$
The asymptotic behaviour given by eq.\asil\ shows that the roots
$\t_k$ accumulate at $\pm\infty$.

The smaller positive root will be called $b \equiv \t_0$.
Then, since  there are no roots in the interval $ -b \leq \t \leq b $, it is
convenient to write eq.\ecro\ as
$$
 \rho(\t) - {{m\b}\o{2\pi}} \cosh\t - \int_{-b}^{+b} d\mu \;
G_0(\t-\mu)\; \rho(\mu) = B(\t)
\eqn\prep
$$
where
$$
 B(\t) \equiv \int_{|\mu|>b} d\mu \;
G_0(\t-\mu)\;\left[\rho(\mu) - \sum_{k\in\ZZ}\delta(\mu - \t_k)\right]
\eqn\betet
$$
For small $\b$, eq.\asil\ applies and we find
$$
b \buildrel {\b\to 0} \over = \log\left({{2\pi}\o{m\b}}\right)\to
+\infty.
$$
Hence, for small $\b$   $ B(\t)$ tends to zero and we can consider
it  a perturbation.

Eq.\prep\ has the appropriate form to be transformed into a
 Riemann-Hilbert problem upon Fourier transformation. Notice that the
function $ {{m\b}\o{2\pi}} \cosh\t -\rho(\t) $ vanishes for large
$\t$ and can be Fourier transformed as follows:
$$\eqalign{
X_+(\om) \equiv & \;e^{+i\om b} \int_{+b}^{+\infty} d\t \;e^{-i\om\t}\left[
  {{m\b}\o{2\pi}} \cosh\t -\rho(\t) \right]  \cr
X_-(\om) \equiv & \;e^{-i\om b} \int_{-\infty}^{-b} d\t \;e^{-i\om\t}\left[
  {{m\b}\o{2\pi}} \cosh\t -\rho(\t) \right] \cr}        \eqn\xfun
$$
where the functions $X_{\pm}(\om)$ are analytic for $\mp \Im\om > 0 $.
In addition,
$$
X_+(\om) = X_-(-\om)
$$
Now, Fourier transforming eq.\prep\ yields
$$
[1-{\tilde G}_0(\om)] \int_{-b}^{+b} e^{-i\om\t}\rho(\t) \;d\t =
 {{m\b}\o{2\pi}} \;{\widetilde{\rm ch}}_b(\om) +\;e^{-i\om b} \,X_+(\om)
+  \; e^{+i\om b}\, X_-(\om) + \; {\tilde B}(\om)
\eqn\trafo
$$
where the function $ {\widetilde{\rm ch}}_b(\om)$ is defined by
$$
 {\widetilde{\rm ch}}_b(\om) \equiv  \int_{-b}^{+b} d\t \;
e^{-i\om\t}\; \cosh\t = {{\sinh(1+i\om)b}\o{1+i\om}} +
 {{\sinh(1-i\om)b}\o{1-i\om}}
\eqn\chos
$$
The kernel $1-{\tilde G}_0(\om)$ can be factorized as follows
$$\eqalign{
1 -  {\tilde G}_0(\om) = & \;{{\sinh({{\pi^2\om}\o{2\g}})}
\o{\sinh({{\pi^2\om}\o{2\g}})+\sinh([{{\pi}\o{2\g}}-1]\pi\om)}}  \cr
= &\; {1 \o {K_+(\om)K_-(\om)}},\cr}
$$
The function $K_{\pm}(\om)$ is analytic and non-zero for $ \mp \Im(\om) > 0$,
$ K_+(\om) = K_-(-\om) $, and explicitly one finds
$$
K_{\pm}(\om) = \sqrt{2\pi} \left({\g\o\pi}\right)^{-i\om/2}
                \left(1-{\g\o\pi}\right)^{1/2-i(\pi/\g-1)\om/2}
        {{\Gamma \left(1-\frac{i\pi\om}{2\g} \right)} \o
         {\Gamma \left(\frac12 - \frac{i\om}2 \right) \,
          \Gamma \left(1 - (\frac\pi\g -1)\frac{i\om}2 \right) }}
$$
Let us define
$$
f_{\pm}(\om)\equiv  e^{\mp i\om b} \int_{-b}^{+b} d\t \;
e^{-i\om\t}\; \rho(\t).
$$
The function $f_{\pm}(\om)$ is analytic in $\mp \Im \om  > 0 $.
Eq.\trafo\ yields a pair of equations
$$\eqalign{
{{f_+}\o{K_+}} = & K_- \left[  e^{-i\om b} \; \frac{m\b}{2\pi}\,
{\widetilde{\rm ch}}_b+ X_+  \;e^{-2 i\om b} + X_- + e^{-i\om b}\; {\tilde B}
\right]   \cr
{{f_-}\o{K_-}} = & K_+ \left[  e^{+i\om b}\;  \frac{m\b}{2\pi}\,
{\widetilde{\rm ch}}_b + X_- \; e^{+2 i\om b} + X_+ + e^{+i\om b}\; {\tilde B}
\right] . \cr } \eqn\prh
$$
Projecting eqs.\prh\ into the half-plane $\Im \om > 0 $ yields
$$\eqalign{
K_-\,  X_- & + \left[K_-  \left( \frac{m\b}{2\pi}\,
{\widetilde{\rm ch}}_b   \,+  {\tilde B} \right)\, e^{-i\om b} \right]_-
+ \left[K_-\, X_+ \, e^{-2 i\om b} \right]_- = 0  \cr
{{f_-}\o{K_-}} \,- &  \left[K_+  \left(\frac{m\b}{2\pi}  \,
{\widetilde{\rm ch}}_b \, +  {\tilde B} \right)\, e^{i\om b}
\right]_- - \left[K_+ \, X_- \, e^{+2 i\om b} \right]_- = 0 . \cr} \eqn\prhd
$$
The projection $[ \ldots ]_{\pm}$ can be explicitly performed as
$$
\left[F(\om)\right]_{\pm} = \pm \int {{d\om^\prime}\o{2\pi i}} \;
{{F(\om^\prime)}\o {\om - \om^\prime}} ~,\quad  \mp \Im \om > 0 ~ .
$$
Equations analogous to  \prhd\ follow upon the exchange $\om
\leftrightarrow -\om , ~ \pm \leftrightarrow \mp $.

More explicitly, the first of eqs.\prhd\ reads
$$\eqalign{
K_-(\om)\, X_-(\om) = &\intf  {{d\om^\prime}\o{2\pi i}} ~ {{ e^{-i\om^\prime
b}}\o
{\om - \om^\prime}}\;K_-(\om^\prime)\;  \left[ {{m\b}\o{2\pi}}  \;
{\widetilde{\rm ch}}_b(\om^\prime) + {\tilde B}(\om^\prime) \right] \cr
& + \intf {{d\om^\prime}\o{2\pi i}} ~ {{ e^{-2 i\om^\prime b}}\o
{\om - \om^\prime}} \;K_-(\om^\prime)\, X_+(\om^\prime) ~,~  \;  \; \Im(\om)>0
\; .\cr}  \eqn\prht
$$
Then, inserting eq.\chos\ in eq.\prht\ and evaluating by residues part of
the terms, yields
$$\eqalign{
K_-(\om)\, X_-(\om) =& {{im\b} \o {4\pi}}\left[ e^b\;
 {{K_-(\om) - K_-(i)}\o{\om -i}} + e^{-b} \;
{{K_-(\om) }\o{\om+i}} \right]  \cr
 & + \intf {{d\om^\prime}\o{2\pi i}} ~ {{ e^{-2 i\om^\prime b}}\o
{\om - \om^\prime}}\; K_-(\om^\prime)\,\left[ X_-(-\om^\prime) -  {{im\b} \o
{4\pi}} \left(
{{e^b}\o{-\om^\prime-i}} + {{e^{-b}}\o{-\om^\prime+i}} \right) \right] \cr
 & +  \intf {{d\om^\prime}\o{2\pi i}} ~ {{ e^{-i\om^\prime b}}\o
{\om - \om^\prime}}\; K_-(\om^\prime){\tilde B}(\om^\prime)\; . \cr}
\eqn\prhc
$$
Setting
$$
v(\om)\equiv  K_-(\om)\,\left[ X_-(\om)- {{im\b} \o {4\pi}}
\left({{e^b}\o{\om-i}} + {{e^{-b}}\o{\om+i}} \right)\right],
\eqn\defv
$$
we see that $v(\om)$ is analytic in $\Im \om > 0$ except for a simple
pole at $\om = i $. We then obtain from eqs.\prhc\ and \defv
$$
v(\om) = -{{i m\b e^b}\o {4\pi}  }
{{K_-(i)}\o{\om-i}} \;+ \intf {{d\om^\prime}\o{2\pi i}} ~
{{ e^{2 i\om^\prime b}}\o {\om + \om^\prime}} \;\a(\om^\prime)\,v(\om^\prime)
  \; +
\intf {{d\om^\prime}\o{2\pi i}} ~ {{ e^{i\om^\prime b}}\o
{\om + \om^\prime}}\; K_+(\om^\prime){\tilde B}(\om^\prime)\; .
\eqn\finrh
$$
where
$$
\a(\om) \equiv {{K_+(\om) }\o{K_-(\om) }}
$$
The ground state energy $E$ can be conveniently expressed in terms of
the residue of $X_-(\om)$ at $\om = -i $.
In fact, using eqs.\gssf\ for $\eta \to 0^+$, \regu , \denu\ and \dpz,
$E$ may be written as
$$
        E = m ~ \intf d\t\; \cosh\t~\left[ \rho(\t) - \sum_{k\in\ZZ}
        \delta(\t - \t_k) \right]
$$
Comparing this with eqs. \ecro\ and \xfun, we obtain
$$
        X_-(\om) \buildrel {\om \to -i} \over \simeq
        {{-iE}\o{m\pi}} \tan{{\pi^2}\o{2\g}}{{e^{-b}}\o{\om+i}}
$$
that is
$$
        E= {{i m\b}\o {\pi}}e^b\, \cot{{\pi^2}\o{2\g}} \;
                 \lim_{\om\to -i} \,(\om + i)  X_-(\om)
$$
We can also express $E$ in terms of $v(-i)$
$$
        E =  m \,\cot{{\pi^2}\o{2\g}}
        \left[ -{{m\b}\o 4} + {{i \pi e^b}\o {\xi}}\;v(-i) \right]
\eqn\enrh
$$
where
$$
\xi \equiv  \lim_{\om\to -i} {{K_-(\om)}\o{\om + i}}
= \; {{\pi}\o{\sqrt{2\g}}}\left(1 - {{\g}\o{\pi}} \right)^{1 -
{{\pi}\o{2\g}}} ~ {{\Gamma\left(1 - {{\pi}\o{2\g}}\right)}\o
{\Gamma\left({3 \o 2} - {{\pi}\o{2\g}}\right)}}
$$
Eqs.\finrh\ and
\enrh\ are exact and provide the starting point to systematically
compute the high temperature (large $b$) behaviour.  Here we limit
ourselves to a qualitative analysis: to start we notice that the first
term in eq.\enrh\ exactly provides minus the bulk free energy, which
supposedly is the only term regular at $m\b=0$. Next we observe that
the integration path in the first integral of eq.\finrh\ can be
deformed into the path $C_+$ which runs around the positive imaginary
axis.  Thus we see that this integral gives poles to $v(\om)$ in the
half--plane $\Im\om<0$ in correspondence with the poles of $K_+(\om)$
for $\Im\om>0$ (at $\om =i$, $K_+(\om)$ has a simple zero).  These
poles are located at $\Im\om=\xi_n\equiv (2\g/\pi)\;n$,
$n=1,2,\ldots\;$, so that the integral contributes a series of terms
proportional to $\, v(i\xi_n)\exp(-2b\, \xi_n) \,$ to $v(-i)$ and hence to the
energy.  Neglecting for the moment the second integral in eq.\finrh,
we  evaluate $v(\om)$ iteratively for large $b$, and conclude
that $v(-i)$ has an asymptotic expansion in powers of $\exp(-2b\xi_1)
\simeq (m\b/2\pi)^{4\g/\pi}$, including the power zero. Inserting this
in eq.\enrh\ we  find that the energy has a term proportional
to $(m\b)^{-1}$ plus a series of terms with $(m\b)^{-1}$ times integer
powers of $(m\b)^{4\g/\pi}$. The effects of the second integral in
eq. \finrh\ can be taken into account by applying the Euler-Maclaurin
formula to the definition of $B(\t)$, eq.\betet. To second order in the
Euler-Maclaurin expansion we then find
$$
        {\hat B}(\om) \simeq \left[
        -\cos\om + {{\om\sin\om b}\o{6\rho(b)}} + \ldots \right]
$$
where $\rho(b)$ can be evaluated to this order by self--consistency.
This leads to changes in the coefficents of the expansion of $v(-i)$
and hence of the energy, but leaves unchanged the analytic structure.
The same would happen by pushing further the Euler-Maclaurin
expansion.
The structure of this expansion and
the powers $(m\b)^{4\g n/\pi}$ agree with Perturbed Conformal Field
Theory, since the perturbing operator (the mass term in the mT model
or the cosine term in the sG model) has exactly the scale dimension
$2(1-\g/\pi)$.

\chapter{Comparison against standard TBA in the IR limit $m\b\to\infty$.}

We can write the non-linear integral equation \ours\ for the massive
Thirring model as
$$
        \nu_f \equiv \e_f - r\cosh\t = -G_0*L_f + G_1*L_{\bar f}  \eqn\nues
$$
where $ r \equiv m \b $.
In Fourier space the two kernels are related by
${\tilde G}_1(k)={\tilde G}_0(k)e^{\pi k}$, so that eq.\nues\ reads there
$$
        {\tilde\nu}_f = {\tilde G}_0\left( -{\tilde L}_f + e^{\pi k}
                {\tilde L}_{\bar f}  \right)                     \eqn\fours
$$
We have to lowest order

$$
        L_f(\t) = \log[1+e^{-\e_f(\t)}] \simeq e^{-r\cosh\t} \equiv g(\t)
$$

and thus, to the same first order in the uniformly small $g$
$$
        {\tilde \nu}_f \simeq {\tilde G}_0 (-1+e^{\pi k}) {\tilde g}
$$
Hence, to second order in $g$
$$\eqalign{
        L_f+L_{\bar f} & \simeq e^{\e_f} -\frac12 e^{-2\e_f} +
                              e^{\e_fb} -\frac12 e^{-2\e_fb} \cr
                &  \simeq     g(2-\nu_f-\nu_fb) - g^2        \cr
                &  = 2g(1-g) + g W\ast g       \cr}             \eqn\ourris
$$
where the kernel of the convolution operator $W$ has Fourier transform
$$
        \til W(k) = 1-4\til G_0(k)\sinh^2\pi k/2 =
        1 - {{2\sinh \left( \frac{\pi^2}{2\g}-1 \right)k \,
          \sinh^2 \frac\pi 2 k } \o
        {\sinh \left( \frac{\pi^2}{2\g}-\frac12 \right)k \,
        \cosh  \frac\pi 2 k }}
$$
The traditional TBA is based on the string hypothesis. According to the work
of Takahashi and Suzuki [\tak], one needs
to write $\g/\pi$ as continued fraction
whose structure will fix the type of string configurations allowed as roots
of the BAE at finite temperature $T=1/\beta$. Standard entropy arguments then
lead to a generically infinite set of coupled nonlinear integral equations
[\tak]. When $\g/\pi$ is a rational number, this set of equations can be
reduced to a finite system. The simplest case is obtained when $\g=\pi/n$,
with $n$ an integer larger than 2. One then deals with the following
system of coupled equations for the real pseudoenergies $\e_j(\t)$,
$j=1,2,\ldots,n$
$$\eqalign{
   &    \e_1(\t) = r\cosh\t - D\ast L_2(\t)     \cr
   &    \e_j = -D\ast (L_{j-1}+L_{j+1}) \;,\quad j=2,3,\ldots,n-3  \cr
   &    \e_{n-2} = D\ast (L_{n-3}+2L_n)         \cr
   &    \e_{n-1} = \e_n = -D\ast L_{n-2}        \cr}           \eqn\dntba
$$
where $L_j=\log(1+e^{-\e_j})$ and $D(\t)=[2\pi\cosh(\t)]^{-1}$.
The structure of this system is that of the
$D_n$ Dynkin diagram.   In the limit $r\to\infty$ the functions $L_j$ all
tend to constants, with $L_1\to 0$ in particular. Since $\til D(0)=1/2$
we then get the system of numerical equations
$$\eqalign{
   &    \e_j = -\frac12 \log\left(1+e^{-\e_{j-1}}\right)
                    \left(1+e^{-\e_{j+1}}\right)\;,\quad j=2,3,\ldots,n-3  \cr
   &    \e_{n-2} = -\frac12 \log\left(1+e^{-\e_{n-3}}\right)
                                \left(1+e^{-\e_n}\right)^2      \cr
   &    \e_{n-1} = \e_n = -\frac12 \log\left(1+e^{-\e_{n-2}}\right)
\cr} \eqn\num
$$
with solution
$$\eqalign{
   &    \e_j = - \log\left( j^2-1 \right) \;,\quad j=1,2,\ldots,n-2  \cr
   &    \e_{n-1} = \e_n = -\log(n-2)     \cr}
$$
We now linearize eqs. \dntba\ around this $r=\infty$ solution, by setting
$$\eqalign{
   &    L_1 = \ell_1 =2g    \cr
   &    L_j = \log j^2 + \ell_j \; ,\quad \e_j = -\log(j^2-1)
         - {{j^2}\o{j^2-1}} \; \ell_j \;,\quad j=2,3,\ldots,n-2   \cr
   &    L_{n-1} = L_n = \log(n-1) + \ell_{n-1} \; ,\quad
        \e_{n-1} = -\log(j^2-1) - {{n-1}\o{n-2}} ~
\ell_{n-1}    \cr}  \eqn\setlin
$$
so that the functions $\ell_j$ fulfill the linear system
$$\eqalign{
   &    {{j^2}\o{j^2-1}} \ell_j = D\ast (\ell_{j-1}+\ell_{j+1})
                        \;,\quad j=2,3,\ldots,n-3   \cr
   &    {{(n-2)^2}\o{(n-2)^2-1}} \ell_{n-2} = D\ast (\ell_{n-3}+2\ell_{n-1})\cr
   &    {{n-1}\o{n-2}}\ell_{n-1} = D\ast \ell_{n-2}  \cr}
\eqn\linsys
$$
Clearly all $\ell_j$ are of order $g$, so that we obtain the order $g^2$
for the function $L_1$ from
$$
        L_1 \simeq e^{-\e_1} -\frac12 e^{-2\e_1} \simeq
                2g(1-g) + g D\ast \ell_2                        \eqn\trad
$$
since $\e_1 \simeq r\cosh\t -\log 2 -D\ast \ell_2$. Agreement with the
result \ourris\ of our  TBA, requires
$$
        D\ast \ell_2 = W\ast g                                  \eqn\requi
$$
as soon as $\g=\pi/n$. This requirement is indeed satisfied: by assuming
$\ell_2$ to be given by eq.\requi\ we can easily calculate in Fourier space
all other $\ell_j$'s
from the first $n-3$ equations of the system \linsys. The last equation of
that system must therefore reduce to an identity, and it does.

\refout

\bye